\documentclass[a4paper,11pt]{article}

\usepackage{jcappub}

\usepackage[utf8]{inputenc}

\usepackage{siunitx}

\usepackage{graphicx}
\usepackage{xcolor}

\usepackage{caption}
\usepackage{subcaption}

\usepackage{natbib}
\def\be{\begin{equation}}
\def\ee{\end{equation}}
\def\bea{\begin{eqnarray}}
\def\eea{\end{eqnarray}}

\def\s{{\rm s}}
\def\cm{{\rm cm}}
\def\kpc{{\rm kpc}}
\def\gev{{\rm GeV}}

\newcommand{\new}[1]{#1}
\newcommand{\renew}[1]{#1}

\title{
Constraining Dark Matter Microphysics with the Annihilation Signal from Subhalos
}
\author[a]{Jack Runburg,}
\author[b]{Eric J.~Baxter,}
\author[a]{Jason Kumar}

\affiliation[a]{\mbox{Department of Physics \& Astronomy, University of Hawai`i, Honolulu, HI 96822, USA}}
\affiliation[b]{\mbox{Institute for Astronomy, University of Hawai`i, 2680 Woodlawn Drive, Honolulu, HI 96822, USA}}

\emailAdd{runburg@hawaii.edu}

\abstract{
In the cold dark matter scenario, galactic dark matter halos are populated with a large number of smaller subhalos. 
Previous work has shown that dark matter annihilations in subhalos can generate a distinctive, non-Poisson signal in the gamma-ray photon counts probability distribution function (PDF).  Here we show that the gamma-ray PDF also carries information about the velocity dependence of the dark matter annihilation cross section.  After calculating the PDF assuming $s$-wave and Sommerfeld-enhanced annihilation, we perform a mock data analysis to illustrate how current and future observations can constrain the microphysics of the dark matter annihilation.  We find that, with current Fermi data, and
assuming a dark matter annihilation cross section roughly 
at the limit of current bounds from annihilation in dwarf 
spheroidal galaxies, one can potentially distinguish the 
non-Poissonian fluctuations expected from dark matter 
annihilation in subhalos from Poisson sources, as well as 
from dark matter models with an incorrect
velocity-dependence. 
\renew{We explore how robust these results are to assumptions about the modeling of the galactic gamma-ray background, but further work is needed to determine the impact of realistic astrophysical source populations on our results}.
We also point out a four-parameter degeneracy between the velocity dependence of the dark matter annihilation, the minimum subhalo mass, the power law index of the subhalo mass function, and the normalization of the dark matter signal.  This degeneracy can be broken with priors from N-body simulations or from observational constraints on the subhalo mass function.}

\arxivnumber{2106.10399}

\begin{document}
\maketitle
\section{Introduction}

A key strategy for probing dark matter models is the 
search for the photons that can be produced when 
dark matter annihilates in an astrophysical body.  
Photons are a particularly interesting search channel 
because they  are relatively easy to detect and because they
point back to the source 
(in contrast, charged particles produced in dark matter annihilations 
will be perturbed by magnetic fields on their passage to Earth). This latter feature has 
been used to focus on regions of the sky which  
are 
known regions of large expected dark matter 
density, such as the Galactic Center and dwarf spheroidal 
galaxies (dSphs).  Such searches are sensitive to the velocity-dependence of the dark matter annihilation (i.e.~its microphysics) in several ways, such as via the angular dependence of the signal \citep[e.g.][]{Boddy:2019wfg} and via the relative amplitude of the signal from different dSphs \citep[e.g.][]{Baxter:2021pzi, lu2018:con, Ando:2021jvn, Feng:2010zp}.

Cold dark matter (CDM) models generically predict the existence of a large population of {\it subhalos} within the main dark matter halo of the Milky Way \citep{Diemand:2006ik}.  The subhalo mass function is expected to rise steeply at small masses, meaning that there are many more small subhalos than large ones.  The exact minimum subhalo mass depends on the particle properties of the dark matter, but masses as low as $1 M_{\oplus}$ and below are allowed in several models \citep[e.g.][]{Green:2005}.  Unlike dark matter annihilation searches aimed at the Galactic Center or dSphs, the precise positions of small galactic subhalos are unknown if they are not massive enough to host stars.  In this paper we consider how the velocity dependence of the dark matter annihilation can be probed by studying the statistics of the unresolved gamma-rays sourced by these subhalos.

The number of photons 
arriving from any particular direction  
due to dark matter annihilation in unresolved substructures will generally be 
drawn from a non-Poissonian 
distribution~\citep{Lee:2008fm}.    
The essential reason is that the
photon flux, integrated over a pixel in the sky, will 
have a non-trivial variance driven by fluctuations in 
the number of high-luminosity subhalos within the pixel.
Although, for a fixed  photon flux, the actual photon count 
is drawn from a Poisson distribution, 
the convolution of this 
Poisson distribution with a flux distribution of non-trivial 
variance is generically non-Poissonian.  In particular, a
fluctuation in the number of high-luminosity subhalos 
(that is, a significant fluctuation in the integrated photon 
flux) could yield a fluctuation in the photon count which 
would be highly unlikely to arise from a Poisson 
distribution.
An important feature we will find is that different models 
for the microphysics of dark matter annihilation (in 
particular, 
$s$-wave annihilation versus Sommerfeld-enhanced 
annihilation) produce different 
non-Poissonian distributions, 
which can be distinguished from each other with sufficient 
data.  


The non-Poissonian distribution of photon counts depends 
on the velocity-dependence of the dark matter 
annihilation 
cross section because the characteristic dark matter 
velocity scale depends on the subhalo parameters (such as mass and size)
in a 
manner which is largely determined by dimensional 
analysis~\citep{Boddy:2019wfg}.  
The subhalo luminosity distribution will thus 
vary with the choice of dark matter microphysics, since these different choices will differently weight the distribution of subhalo parameters.
These differing 
subhalo luminosity distributions will lead to different non-Poisson 
photon count distributions, reflecting fluctuations in the photon 
counts resulting from fluctuations in the number of subhalos along 
the line of sight.

\new{
The photon count distribution has previously been studied for the purpose of 
determining if $s$-wave dark matter annihilation in unresolved subhalos 
can be distinguished from 
astrophysical backgrounds \citep{Lee:2008fm,Baxter:2010fr,Somalwar:2020awt}.  
In this work, we will consider if one can distinguish 
between the different non-Poisson count distributions 
which would arise from different choices for the 
dark matter annihilation cross section velocity-dependence.  To this end, we will compare 
the likelihoods which arise from different models for 
dark matter annihilation. 
}

Our 
strategy will be to estimate the subhalo parameter distribution 
using scaling relations obtained from numerical simulations.  
From this, we will be able to obtain the non-Poissonian 
photon count distribution arising from $s$-wave and 
from Sommerfeld-enhanced (in the Coulomb limit) dark 
matter annihilation.

Recently,  
the photon counts distribution has also
been used to investigate 
whether the GeV photon excess from the Galactic Center (GC) results from the annihilation 
of a smooth dark matter component, 
or from a population of 
unresolved pulsars (see, for 
example,~\cite{Lee:2015fea, Bartels:2015aea,Leane:2019xiy,Leane:2020nmi,Leane:2020pfc,Buschmann:2020adf, calore_dissecting_2021}).  
The GC analyses in particular point to a message we will echo: 
distinguishing a signal from 
background based on the non-Poissonian nature of the 
distribution observed in data can only be useful if one 
has some handle on what the signal and background 
distributions actually are.

We will see that if the background distributions are 
well-modeled, then an experiment which is able to detect 
a photon excess above background would also likely be 
able to determine if this excess is more consistent with 
$s$-wave 
dark matter annihilation, 
Sommerfeld-enhanced dark matter annihilation, or 
a Poisson source.
But, as with analyses of the 
GC, we will find 
that it can be more difficult to distinguish 
between models for the source distribution if the 
background is mismodeled.  But some discrimination 
power remains, even in the presence of mismodeled 
backgrounds, because we are not just looking for 
non-Poisson fluctuations, but are comparing the 
likelihoods arising from particular non-Poissonian 
distributions.  

\renew{We note that our treatment of astrophysical backgrounds ignores the presence of unresolved non-subhalo gamma-ray point sources, such as blazars.  While our analysis establishes that information about the dark matter annihilation velocity dependence is encoded in the photon counts distribution, and that this information is sufficient to place useful constraints in the presence of some backgrounds, we postpone a complete treatment of all possible astrophysical backgrounds to future work.  Depending on the degeneracy between the astrophysical backgrounds and the dark matter signal, some degradation of the dark matter constraints may result.}

As a further complication, we will find that 
the differences in photon count distributions arising from 
different dark matter microphysics models can also be replicated 
by drastic changes to the distribution of subhalo parameters.  As 
background modeling improves, and as the subhalo distributions 
become more tightly constrained by theory \citep[e.g][]{Madau:2008}, and data \citep[e.g][]{Xu:2009,Vegetti:2010,Ostdiek:2020},
our results will become more 
robust.

The plan of this paper is as follows.  In \S\ref{sec:subhalo_pdf} we describe the basic formalism for predicting the photon counts distribution from galactic subhalos; in \S\ref{sec:vel_dependence} we describe how this formalism can be modified to include the impact of velocity-dependent dark matter annihilation; in \S\ref{sec:background_model} we describe our model for gamma-rays produced by astrophysical (i.e.~non-dark matter) sources; in \S\ref{sec:simulated_data} we describe the generation and analysis of simulated data.  We present our main results in \S\ref{sec:results} and conclude in \S\ref{sec:conclusion}.

\section{The Photon Counts Distribution from Subhalos}
\label{sec:subhalo_pdf}

We will consider an analysis in which we ignore the photon energy information, beyond requiring \renew{the photon energies} to be within the detector acceptance 
window.   \new{While energy information could in principle improve constraints on the dark matter properties,}\renew{ a full analysis of the energy-dependent photon counts distribution is highly nontrivial \citep{EPDFABC}}.  \renew{Previous analyses have considered the energy information and PDF information separately, for instance by analyzing the PDF in bins of energy \cite{Lisanti:2016,Zechlin:2016}.  } \new{By ignoring energy information, we make it feasible to calculate the exact likelihood of the  photon counts.  Furthermore, this choice also makes our analysis less sensitive to the precise spectrum of annihilation radiation}, \renew{meaning that our results apply across a broad range of dark matter models}.  \new{Our main conclusion --- that the photon counts distribution contains information about dark matter microphysics --- is not impacted by the exclusion of energy information. }  
We will assume that the sky is divided into 
pixels, and consider only the number of photons seen by the detector in each 
pixel over a given exposure \renew{(i.e. summing over all photon energies within the acceptance window)}.  \renew{We note that in \cite{EPDFABC}, it was found that the inclusion of energy information into a PDF analysis similar to that considered here could improve constraints on dark matter parameters by roughly a factor of two for some models, so the present analysis may be considered conservative in that sense.}

The basic quantity which we wish to determine 
is $P_C (C_i)$, 
the probability of observing $C_i$ counts in the $i$th 
pixel over some observation time.  To determine this probability distribution function, we 
follow the analysis of~\cite{Baxter:2010fr}.  We then find
\bea
\label{eq:p_of_c}
P_C(C_i) &=& \int dF~ P_{sh}(F; \psi_i) ~ 
{\cal P}[C_i ; E_i (F + F_i^{bgd}) ] , 
\eea
where $\psi_i$ is the angle between the $i$th pixel and the Galactic 
Center. $P_{sh} (F; \psi_i)$ is the probability distribution 
for a flux $F$ of photons 
(with energy within the acceptance of the instrument) 
to be produced by 
dark matter annihilation within all subhalos located in the $i$th pixel. 
$F_i^{bgd}$ is the  
flux of photons due to 
astrophysical foregrounds within the energy 
range of the instrument, integrated over the solid angle of the $i$th pixel.\footnote{Note, we are ignoring the photon 
flux arising from dark matter annihilation in the smooth 
component of the galaxy halo.  But at the latitudes on 
which we will focus, the contribution from the smooth 
component will be 
subdominant~\citep{Springel:2008by}.}
$E_i$ is the exposure of the instrument to the $i$th pixel (i.e. $E_i$ has units of ${\rm area} \times {\rm time}$), and 
$E_i (F + F_i^{bgd})$ is thus the expected number of photons, from all sources, arriving 
from the $i$th pixel.  ${\cal P}[ C_i; E_i (F + F_i^{bgd}) ]$ is then the Poisson
probability for obtaining $C_i$ counts from a Poisson distribution with mean given by 
$E_i (F + F_i^{bgd})$. 

To find $P(C_i)$, we must determine $P_{sh} (F; \psi_i)$.  
This probability distribution is in turn determined 
by $P_1 (F; \psi_i)$, which is defined as the probability 
distribution for a single subhalo at angle 
$\psi_i$ to produce a photon flux $F$ via dark matter 
annihilation.  Essentially, $P_{sh} (F; \psi_i)$ is 
given by the product of the probability of there being 
$m$ subhalos within the $i$th pixel and the 
probability that those $m$ subhalos produce a total 
flux $F$ of photons, marginalized over all $m$.  We assume that 
the probability of there being $m$ subhalos in the 
$i$th pixel is Poisson distributed about a mean 
value $\mu (\psi_i)$.  
In that case, we may express $P_{sh}$ as~\citep{Lee:2008fm} 
\bea
\label{eq:p1_to_pf}
P_{sh} (F; \psi_i) &=& 
{\cal F}^{-1} \left\{ \exp \left[ 
\mu(\psi_i) \left( {\cal F} \{ P_1 (F; \psi_i)  \} -1  
\right)
\right] \right\} ,
\eea
where ${\cal F}$ denotes the Fourier transform with respect 
to $F$ (and ${\cal F}^{-1}$ is the inverse Fourier transform), with normalizations defined as in \cite{Lee:2008fm}.  Given the Poisson statistics of $m$, the expected standard deviation of the number of subhalos along a line of sight is $\sqrt{\mu}$, so as $\mu$ becomes larger, the distribution of $m$ will become narrower around $\mu$.  The result is that the flux distribution from all subhalos, $P_{sh}(F)$, will become more $\delta$-function-like as $\mu$ increases, leading to a photon count distribution $P_C$ which is more Poisson-like.

$P_1 (F; \psi_i)$ is then defined in terms of 
$P_L (L_{sh}; \ell, \psi_i)$, the probability distribution 
for a subhalo located at distance $\ell$ along the line of 
sight at angle $\psi_i$ from the GC to have intrinsic luminosity $L_{sh}$.  In particular, one finds
\bea
\label{eq:p1}
P_1 (F; \psi_i) &\propto& \theta (F_{\rm max} - F) 
\int_0^{\ell_{\rm max}} d\ell \int dL_{sh}~ 
P_L (L_{sh}; \ell, \psi_i)~ 
\delta \left(F - \frac{L_{sh}}{4\pi \ell^2} \right) ,
\eea
where $\ell_{\rm max}$ \new{is the maximum distance of a subhalo from the observer, calculated at a given angle $\psi_i$.  We discuss the assumed subhalo distribution in more detail below.  The step function, $\theta$, enforces an upper limit to the flux from a single object, which arises because it is assumed that halos which produce a flux larger than 
$F_{\rm max}$ would be resolved as point sources and not considered part of the unresolved photon flux.  }\renew{Although the resolved source flux limit of Fermi depends on energy and spatial position, for simplicity we adopt a reasonable value of $F_{\rm max}=10^{-2}\,\,\mathrm{photons}\,\,\mathrm{cm}^{-2}\,\,\mathrm{yr}^{-1}$, which corresponds to the median flux over the energy range of 1 to 100 GeV in the \textit{Fermi} LAT 12-year point source catalog \cite{Fermiptsource}.
In practice, the precise value of $F_{\rm max}$ assumed here is not important, since it is many orders of magnitude larger than the regime relevant for our analysis (see Fig.~\ref{fig:p1}).    Lowering $F_{\rm max}$ by an order of magnitude to  $F_{\rm max}=10^{-3}\,\,\mathrm{photons}\,\,\mathrm{cm}^{-3}\,\,\mathrm{yr}^{-1}$  would correspond to a flux limit lower than 99.5\% of sources in the 12-year point source catalog, and would have no impact on our results.  Increasing $F_{\rm max}$ by an order of magnitude also has no impact on our results.}

We can proceed further by relating the probability distribution for the subhalo luminosity to the subhalo mass function and the conditional luminosity function, 
$P_L [L_{sh} | M, r (\ell, \psi_i)]$,  
which gives the
probability for a subhalo of mass $M$ at distance  
$r$ with respect to the GC to have luminosity 
$L_{sh}$.  In particular, one finds
\bea
P_L (L_{sh}; \ell, \psi_i) d\ell &\propto& 
\ell^2 d\ell \int_{M_{\rm min}}^{M_{\rm max}} dM~
\frac{d^2 N (r(\ell, \psi_i))}{dM dV} 
\times P_L [L_{sh} | M, r (\ell, \psi_i)] ,
\eea
where $d^2N / dM dV$ is the subhalo mass function, 
and $M_{\rm min}$ and $M_{\rm max}$ are the minimum and 
maximum subhalo masses, respectively.

For the subhalo mass function, we adopt a form 
drawn from numerical simulations:
\bea
\label{eq:mass_func}
\frac{d^2 N(r)}{dM dV} &=& A \frac{(M/M_\odot)^{-\beta}}
{\tilde r (1+\tilde r)^2} ,
\eea
where $\tilde r = r / 21~\kpc$, 
$A = 1.2 \times 10^4 \, M_\odot^{-1} \kpc^{-3}$, 
and $\beta = 1.9$~\citep{Springel:2008cc}.  \renew{In Eq.~\ref{eq:mass_func}, we have assumed that the subhalo radial distribution follows the total dark matter distribution in the galaxy, which is in turn assumed to follow a Navarro-Frenk-White (NFW) profile \citep{NFW}}.\footnote{Note that Milky Way dark matter 
distribution need not be described by an NFW profile (for a review of this issue, see~\cite{Salucci:2018hqu}).  \renew{A different distribution of dark matter within the galaxy would change the variation in the mean annihilation flux across the sky, but would not change the fundamental non-Poisson nature of the signal from dark matter subhalos, which is the main focus of our analysis.}.} \renew{When subhalos in simulations are selected on quantities that are robust to tidal evolution (e.g. mass prior to accretion), the subhalo distribution is indeed found to closely track the total dark matter distribution \cite{Kravtsov:2010}.  Since tidal disruption has a comparatively small impact on the annihilation rate compared to its impact on the total mass, and since we are primarily interested in small subhalos for which the impact of dynamical friction is small, the adopted prescription for the subhalo distribution is reasonable.}   Here, $21~\kpc$ is roughly the scale radius of the Milky Way, assuming its dark matter distribution is described by a   \new{Assuming a halo concentration of $c \sim 10$ for the Milky Way \cite{Diemer:2015}, this leads to a virial radius of roughly $210~\kpc$, which we take as the maximum halo radius when calculating $\ell_{\rm max}$ in Eq.~\ref{eq:p1}.}
Note that $P_{sh}(F ; \psi_i)$ 
is largely insensitive to $\ell_{\rm max}$.  
Although the number of subhalos along the line 
of sight grows logarithmically with 
$\ell_{\rm max}$, the photon flux attributable to 
most distant subhalos scales as 
$\ell_{\rm max}^{-2}$, implying that the most 
distant subhalos collectively provide a 
negligible contribution to $P_{sh}$.

Our next task is to determine 
$ P_L [L_{sh} | M, r (\ell, \psi_i)] $.  
Following the approach of 
\cite{Koushiappas:2010fk}, 
we choose
\bea
P_L [\ln L_{sh} | M, r (\ell, \psi_i)] &=& 
\frac{1}{\sqrt{2\pi} \sigma} \exp \left[- 
\frac{(\ln L_{sh} - \langle \ln L_{sh} \rangle)^2}{2\sigma^2} 
\right] ,
\eea
where $\langle L_{sh} \rangle$ is the mean luminosity for a subhalo 
of mass $M$ at distance $r$ from the GC, and we 
take $\sigma = 0.74$.\footnote{There is 
a mild dependence of the variance on $r$ and the halo mass, 
but, for simplicity, we ignore this henceforth.  \renew{We have confirmed that ignoring this dependence has a negligible impact on our analysis.}}

In \cite{Koushiappas:2010fk} the mean subhalo 
luminosity in galactic-scale halos was estimated from numerical simulation results 
assuming $s$-wave annihilation.  
Adopting model $C_0$ of \cite{Koushiappas:2010fk}, which corresponds roughly to Milky Way-like galactic halos, one finds for $s$-wave annihilation
\bea
\label{eq:subhalo_luminosity}
\langle \ln \left( L_{sh}
/ \s^{-1} \right) \rangle &=& 
77.4 + 0.87 \ln (M / 10^5 M_\odot) 
-0.23 \ln (r / 50\kpc)
+ \ln \left(\frac{8\pi \Phi_{PP}}{10^{-28} \cm^3 \s^{-1} 
\gev^{-2}} \right) ,  \nonumber \\
\label{eq:MeanLuminosity}
\eea
where $\Phi_{PP}$ is a normalization parameter which is 
determined by the dark matter microphysics, as we will see in the next section.  \renew{Recently,  \cite{Grand:2021} has highlighted the possibility that baryonic effects can lead to enhanced tidal disruption of subhalos, leading to a suppression in their annihilation luminosities, particularly for subhalos on orbits with small pericenters.  It remains somewhat unclear, however, the extent to which these effects impact the very small subhalos most relevant to the present analysis.  Regardless of the level of baryonic suppression, we expect the main effect considered in this work --- namely, the impact of the annihilation velocity dependence on the photon counts PDF --- to still hold true.  We postpone a full investigation of baryonic effects on the PDF to future work. }

We will now generalize 
the expression for $\langle \ln L_{sh} \rangle$ to the scenario in which 
dark matter annihilation 
has a non-trivial velocity-dependence.
To do so, 
we must consider the generic dependence of the subhalo luminosity on 
the subhalo parameters.

\section{Incorporating the impact of velocity-dependent annihilation}
\label{sec:vel_dependence}

We assume that dark matter is a 
\new{self-conjugate}
particle which 
\new{is its own anti-particle and has an 
annihilation}
cross section that can 
be written as $\sigma v = (\sigma v)_0 \times S(v/c)$, where 
$(\sigma v)_0$ is independent of the relative velocity 
$v$, and $S(v/c) = (v/c)^n$.  The standard case of $s$-wave annihilation thus corresponds to $n=0$, while $p$-wave and $d$-wave annihilation correspond to $n=2$ and $n=4$, respectively.
Sommerfeld annihilation in the Coulomb limit corresponds to $n=-1$.

The flux of photons arising from dark matter annihilation in 
a subhalo can be written as 
\bea
\Phi &=& \Phi_{PP} \times \bar J, 
\eea
where 
\bea
\Phi_{PP} &=&  \frac{(\sigma v)_0}{8\pi m_X^2} \bar N_\gamma ,
\eea
and $\bar J$ is the so-called $J$-factor, integrated over the full 
extent of the subhalo.  Here, $m_X$ is the mass of the 
dark matter particle, and $\bar N_\gamma$ 
is the average number of photons produced per 
annihilation within the energy range of the observatory
\new{($\bar N_\gamma$ will thus depend on the details of the particle physics model, including the annihilation 
channel)}.  
\renew{In particular, $\bar N_\gamma$ depends on the branching 
fraction for annihilation to any particular final state, as well
as the photon spectrum produced when any of the particles in 
that final state decay.  For any model, $\bar N_\gamma$ can be 
determined using available numerical packages~\cite{Sjostrand:2007gs,Cirelli:2010xx,Reimitz:2021wcq}.  }

If we assume that the subhalo is at a 
distance $D$ from the observatory which is much larger than 
the size of the subhalo, then we may approximate the
integrated $J$-factor as
\bea
\bar J &=& (1/D^2) \int d^3 r \int d^3 v_1 \int d^3 v_2~ 
f (\vec{r}, \vec{v}_1) f (\vec{r}, \vec{v}_2) 
\times (v/c)^n ,
\eea
where $\vec{v} = \vec{v}_1 - \vec{v}_2$ is the 
relative velocity, and $f(\vec{r},\vec{v})$ is the 
dark matter velocity distribution~\citep{Robertson:2009bh,Belotsky:2014doa,Ferrer:2013cla,Boddy:2017vpe,Zhao:2017dln,Petac:2018gue,Boddy:2018ike,Lacroix:2018qqh,Boddy:2019wfg,Boddy:2020,Ando:2021jvn}.

One key assumption we will make is that all dark matter 
subhalos have the same \new{functional form of their density profiles $\rho(r)\equiv\rho_s\tilde{\rho}(r/r_s)$} and are parameterized by two 
dimensionful parameters: the scale density $\rho_s$ and the 
scale radius $r_s$.
\new{
In general, subhalos are also characterized by a tidal radius $r_t$, outside 
of which dark matter is tidally stripped from the subhalo.  The tidal radius depends on the subhalo density profile, as well as its orbit through the parent halo of the Mikly Way.  In general, though, it appears that for the smallest subhalos, $r_t$ is typically larger than $r_s$, and the impact of $r_t$ on the $J$-factor is fairly small (i.e. less than an order of magnitude) \cite{Delos:2019}.
For the halo profiles which are typically considered, 
the dark matter density falls off steeply with $r$ 
outside the scale radius; for example, for an NFW profile, $\rho (r) \propto r^{-3}$ for $r \gg r_s$.  Thus, although the mass of the subhalo grows logarithmically with $r$, 
the annihilation rate (which scales as $\rho^2$) becomes 
negligible at large distances.}

\new{If the only dimensionful parameters of the subhalo 
are $\rho_s$ and $r_s$,} the dependence of 
$\bar J$ on the halo parameters is determined entirely 
by dimensional analysis.  Since the only velocity scale 
in the problem is $(4\pi G_N \rho_s r_s^2)^{1/2}$, the 
integrated $J$-factor may be written as~\citep{Boddy:2019wfg}
\bea
\bar J &\propto& 
\frac{\rho_s^2 r_s^3}{D^2} 
(4\pi G_N \rho_s r_s^2)^{n/2} , 
\label{eqn:JfactorDependence}
\eea
where the dimensionless proportionality constant is 
determined only by $n$ and the 
\renew{functional}
form of the dark matter 
distribution, but is independent of the parameters 
$\rho_s$ and $r_s$.

Rather than deal with the parameter $\rho_s$, 
it will be convenient to define a scale mass parameter, 
$M_s \equiv \rho_s r_s^3$.  
For an NFW profile,
the halo mass is proportional to $M_s$ up to a logartihmic dependence 
on the tidal radius.
We then see that the photon luminosity of any subhalo can 
be expressed as
\bea
L_n &\propto& \frac{M_s^2}{r_s^3} (4\pi G_N M_s / r_s)^{n/2} ,
\label{eqn:Ln_Dependence}
\eea
up to 
a constant which depends on the functional form of the profile but not on the parameters.

We are still left with two halo parameters, $M_s$ and $r_s$.  However, a 
variety of simulation and semi-analytic results 
have led to a 
scaling relation  
which relates $r_s$ to the halo mass and the distance $r$ from 
the GC \citep{Koushiappas:2010fk}.  In particular, 
the mean luminosity relationship found in 
Eq.~\ref{eq:MeanLuminosity} \new{(originally from \cite{Koushiappas:2010fk})}, 
\bea
\langle L_{sh}(n=0) \rangle \propto M^{0.87} r^{-0.23} ,
\label{eqn:LuminosityRelation}
\eea
implies the scaling relation
\bea
r_s \propto M_s^{0.38} r^{0.08} ,
\eea
\new{where we have assumed $M \propto M_s$ and used Eq.~\ref{eqn:JfactorDependence}}.
This yields a mean subhalo luminosity which scales as 
\bea
\label{eq:param_dep}
\langle L_{sh}(n) \rangle &\propto& M^{0.87} r^{-0.23} 
\times \left(M^{0.62} r^{-0.08} \right)^{n/2} .
\eea
We will adopt this relation in our analysis below.
Note that the proportionality constant, 
which we have omitted in Eq.~\ref{eq:param_dep}, can 
be simply absorbed into the definition of the normalization 
parameter, as in Eq.~\ref{eq:MeanLuminosity}.

Above, we have assumed that dark matter subhalos can be described by a two parameter model such as the NFW profile.  It is known from N-body simulations, however, that dark matter halos can be triaxial objects \citep[e.g.][]{Jing:2002}.  Triaxiality could in principle have an effect on the subhalo $J$-factors.  
\new{However, we note that subhalo triaxiality is found to decrease with decreasing subhalo mass \citep{Vera-Ciro:2014}, suggesting that the very small subhalos considered here may not be significantly triaxial.  Furthermore, it seems unlikely that triaxiality would significantly impact the dependence of the $J$ factors on $n$}.
Since our primary intent is to point out the general impact of velocity-dependent dark matter annihilation on the photon statistics from dark matter annihilation in subhalos, we will ignore subhalo triaxiality below.  \new{We discuss further parameter degeneracies in \S\ref{sec:degeneracy}.}

We now consider how the velocity dependence of the dark matter annihilation impacts the photon counts distribution.  
We first discuss the high flux tail of $P_1(F)$, as this is what sets the non-Poisson tail of $P_{sh}(F)$ that makes it possible to distinguish between the dark matter subhalo signal and smooth backgrounds. 
As described in \cite{Lee:2008fm}, the high-$F$ tail of $P_{sh}(F)$ will follow that of $P_1(F)$ since in the high flux limit, single bright sources are the dominant source of flux.  
Ignoring the weak dependence of $L_{sh}$ on $r$, from Eq.~\ref{eq:param_dep} we can write $L_{sh} \propto M^\alpha$ with $\alpha = 0.87 + 0.31n$.   At large $F$, $P_1 (F) \propto F^\gamma$, with $\gamma = [(1-\beta)/\alpha]-1$ and where $\beta$ is the power law index of the mass function, as in Eq.~\ref{eq:mass_func} \citep{Baxter:2010fr}.  Substituting, we have
\begin{equation}
\label{eq:gamma}
\gamma \equiv \frac{1-\beta}{\alpha}-1 = - \frac{1.03}{1 +0.36n} -1 .
\end{equation}
Consequently, larger $n$ (e.g., $p$-wave or $d$-wave annihilation)
will yield $P_{1}(F)$ that are flatter than the $P_1(F)$ obtained for small $n$ (Sommerfeld or $s$-wave).  Note that the variance of 
$P_1 (F ; n)$ is controlled by $F_{\rm max}$ for $\gamma > -3$, which encompasses every model that we consider.

Below, we will focus on two scenarios of dark matter annihilation: 
$s$-wave annihilation ($n=0$), and 
Sommerfeld-enhanced annihilation~\citep{ArkaniHamed:2008qn,Feng:2010zp} 
in the Coulomb limit ($n=-1$).  $s$-wave annihilation is 
the standard case which is most often considered.  
Sommerfeld enhancement arises if dark matter particles 
have an attractive self-interaction mediated by a light 
particle.  The Sommerfeld-enhanced  scenario is particularly interesting in the case of annihilation in a subhalo because 
the relative velocities of particles bound to a subhalo 
tend to be smaller \new{by about an order of magnitude} than the relative velocities of particles 
bound to the galactic halo \citep{Baxter:2021pzi}.  Thus, the signal from 
Sommerfeld-enhanced dark matter annihilation in a subhalo 
will be enhanced relative to the signal from the GC.  
Conversely, $p$-/$d$-wave annihilation will 
yield signals from subhalos which are suppressed with 
respect to the GC, so we will not consider them further in 
this work.  But although we focus on $s$-wave and Sommerfeld-enhanced annihilation, the general arguments that we make below should also apply to e.g. $p$-wave and $d$-wave annihilation.

\begin{figure}
    \centering
    \begin{subfigure}[t]{0.5\textwidth}
    \centering
    \includegraphics[width=\textwidth]{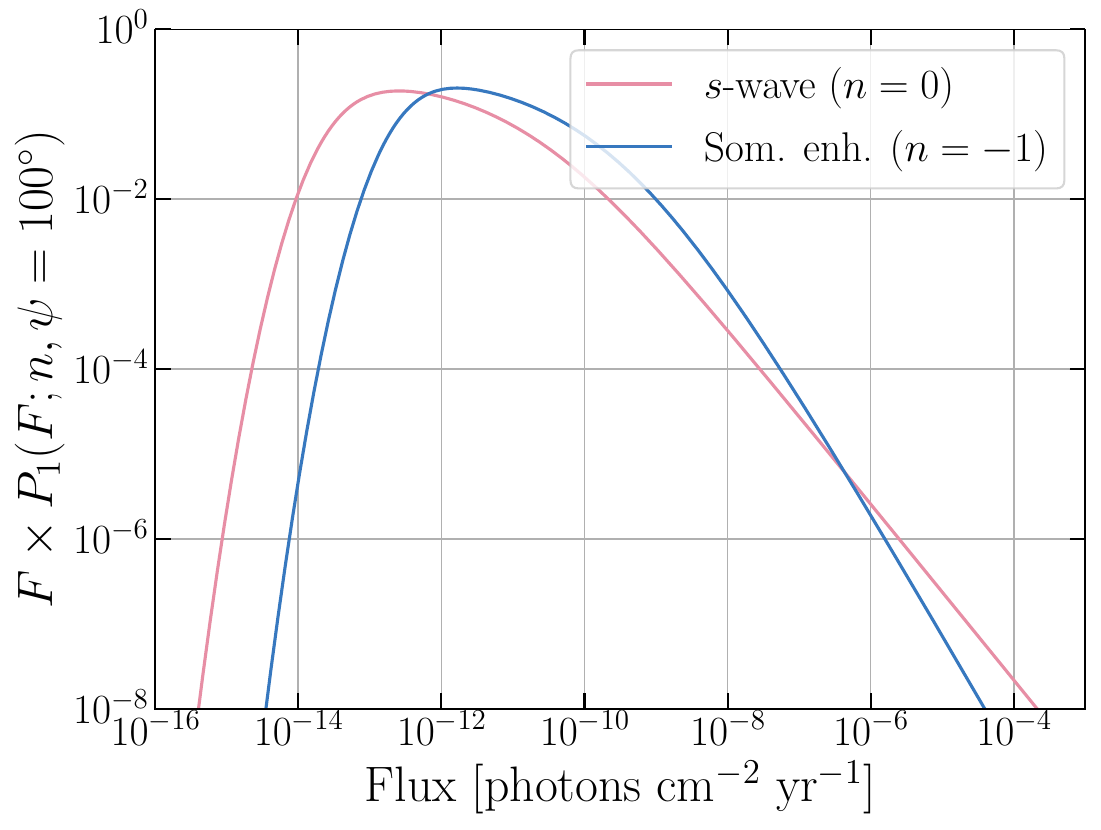}
    \caption{}
    \label{fig:p1}
    \end{subfigure}%
    \begin{subfigure}[t]{0.5\textwidth}
    \centering
    \includegraphics[width=\textwidth]{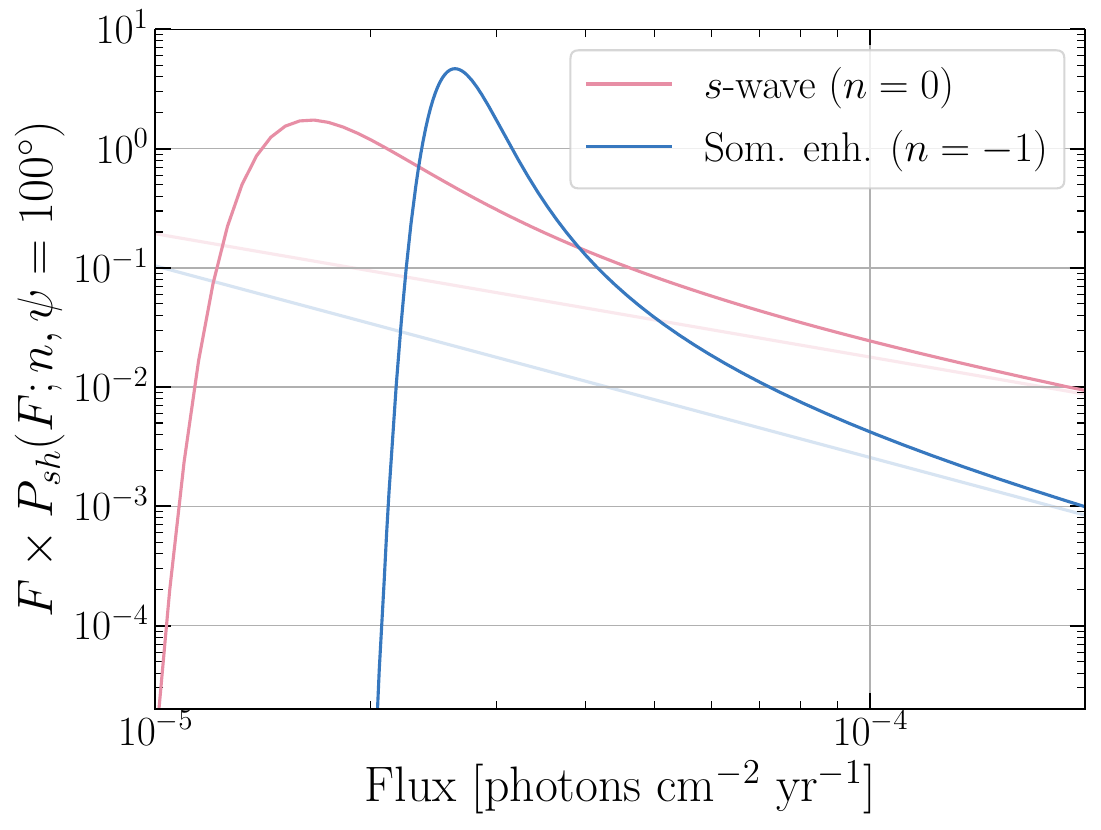}
    \caption{}
    \label{fig:psh}
    \end{subfigure}
    \caption{$(a)$ The DM annihilation flux probability distribution for one subhalo $P_1(F;n,\psi_i$) for $s$-wave ($n=0$) and Sommerfeld annihilation ($n=-1$), 
    at $\psi_0 = 100^\circ$. 
    $s$-wave annihilation results in a flatter power law tail.
    $(b)$ The DM annihilation flux probability distribution from all subhalos along a line of sight, $P_{sh}(F;n,\psi_i$).  
    Because Sommerfeld-enhanced annihilation leads to a larger flux contribution from smaller subhalos (which are much more numerous than large subhalos), $P_{sh}(F;n=-1)$ is more tightly peaked than 
    $P_{sh}(F;n=0)$ (see discussion in \S\ref{sec:subhalo_pdf}).  The faded lines indicate the expected high flux power law tail expected from Eq.~\ref{eq:gamma}. }
    \label{fig:p1psh}
\end{figure}

In Figure~\ref{fig:p1psh}, we plot $P_1 (F ; n, \psi_i)$ (left panel) and 
$P_{sh} (F; n, \psi_i )$ (right panel) 
for $n=0,-1$, 
with $\psi_i = 100^\circ$.
We take $M_{\rm min} = 0.1\,\,\mathrm{M_\odot}$, $M_{\rm max} = 10^{10}\,\,\mathrm{M_\odot}$, yielding 
$\mu(\psi_i=100^\circ) = 68539.8$.
The normalization of the dark matter signal is chosen to be $\Phi_{PP}=  7 \times 10^{-30} \cm^3~\s^{-1} 
~\gev^{-2}$ 
in the case of $s$-wave annihilation, which roughly corresponds to the 
upper limit obtained 
from a search of dwarf spheroidal galaxies, as determined 
by \texttt{MADHAT}~\citep{Boddy:2019kuw}, assuming a photon 
energy range of 1-100~\gev.\footnote{\renew{These bounds are obtained by estimating the expected photon count from each dSph due to ordinary astrophysical processes by counting the number of photons arriving from  similar-sized regions slightly off-axis from each dSph, with 
point sources masked.  This background probability distribution, along with the observed photon count from each dSph, leads to a statistical bound on expected number of photons attributable to dark matter annihilation.  Given the $J$-factors, this in turn leads to a bound on $\Phi_{PP}$.}}
\renew{For 
example, for dark matter with mass $m_X = 100~\gev$ which 
annihilates to $\bar b b$, one would find $\bar N_\gamma 
\sim 13.6$~\cite{Cirelli:2010xx}.  If the $s$-wave annihilation cross section is $(\sigma v)_0 = 1.35 \times 10^{-25} \cm^3 / \s$, then this value of the $\Phi_{PP}$ normalization is realized.}
For the case of Sommerfeld enhanced annihilation, we choose the normalization to be such that the expected number of photons from dark matter annihilation is the same as in the $s$-wave case.
\renew{Note that there is no prediction for the normalization 
in the case of Sommerfeld-enhanced annihilation, because the 
normalization is model-dependent, and any model with 
Sommerfeld-enhanced annihilation is necessarily different 
from a model with $s$-wave annihilation.  But the choice of 
normalization which we make in Figure~\ref{fig:p1psh} 
(both $s$-wave and 
Sommerfeld-enhanced annihilation producing the same number 
of expected photons per pixel) is the most interesting case.  
If two models yield expected photon counts which are very 
different, then it is easy to determine which model is preferred by the data.  The most interesting case, for our purpose, is thus when both models yield the same number of 
expected photons, but differ in the non-Poisson shapes of the 
distributions.}

\begin{figure}
    \centering
    \includegraphics[width=0.6\textwidth]{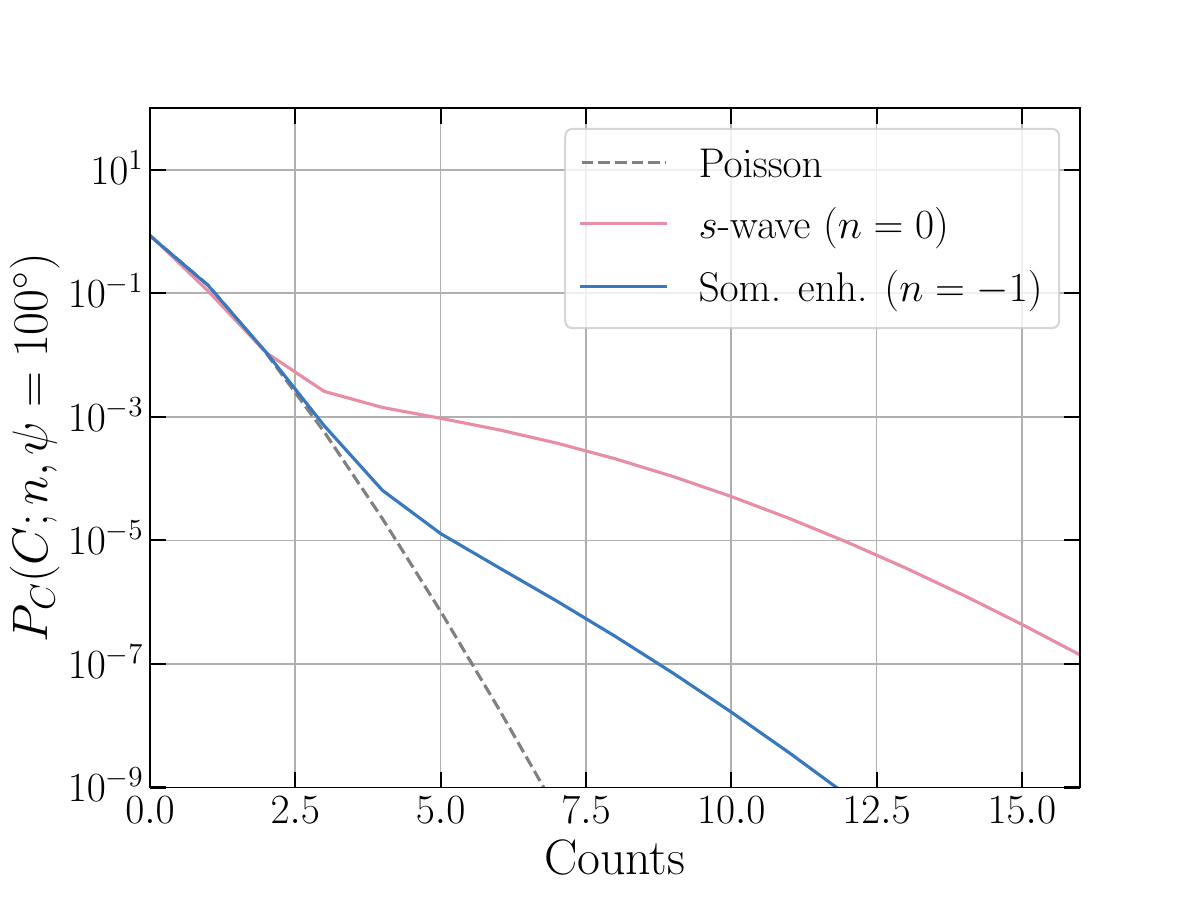}
    \caption{$P_C(C)$ for $s$-wave and Sommerfeld-enhanced annihilation. The dashed line indicates a Poisson distribution with the mean of $P_{sh}(F; n)$. For both $n=0,-1$, there is a higher probability to obtain larger counts compared to the Poisson distribution.  All count distributions have the same mean.}
    \label{fig:pc}
\end{figure}

In Fig.~\ref{fig:pc}, we plot $P_C(C; n,\psi_i)$ for $n=0,-1$ 
(pink and blue solid lines, respectively), setting  $\psi_i=100^\circ$ 
and using the same parameter values as above.  To compute the photon counts distribution we must assume an exposure (see e.g. Eq.~\ref{eq:p_of_c}).   We assume Fermi-like observations with a collecting area of $2000\,{\rm cm}^2$, a field of view of seven sterradians, and an observations time of five years, yielding
an exposure of $5570.5\,\, \mathrm{cm^2\,yr}$. 
We use a pixel size of $0.21\,\mathrm{deg^2}$, corresponding to a \texttt{healpix} map with $N_{\rm side} = 128$.
We also plot a Poisson distribution (dashed, gray line) with mean identical to that of the $P_C(C; n)$ curves.  
Note that $P_C(C; n=-1)$ is much closer to a Poisson distribution than $P_C(C; n=0)$.  
The reason is because 
$P_{sh}(F;n=-1)$ has a high-$F$ tail which falls off 
more steeply than $P_{sh}(F;n=0)$ (see eq.~\ref{eq:gamma} 
and Fig.~\ref{fig:p1psh}), 
implying that $P_{sh}$ has a smaller variance for  the
Sommerfeld-enhanced case.  As $P_{sh}(F)$ begins to look 
more like a $\delta$-function, $P_C(C)$ begins to look more like a Poisson distribution.

\section{Background Model}
\label{sec:background_model}

To perform a realistic mock data analysis, we must include the impact of astrophysical gamma-ray backgrounds.  Our estimate of astrophysical backgrounds is derived from the models developed by \cite{collaboration2020}.  
These models include a diffuse galactic component (\texttt{gll\_iem\_v07.fits}) and an isotropic component (\texttt{iso\_P8R3\_SOURCE\_V2\_FRONT\_v1.txt}).
For simplicity, we use the front-converting events in the energy range $1\,\mathrm{GeV}\,-\,3\,\mathrm{TeV}$ because the acceptance is roughly constant for these events \new{with an effective area of $\sim 2000 \,\mathrm{cm}^2$}.\footnote{\url{https://www.slac.stanford.edu/exp/glast/groups/canda/lat_Performance.htm}}
The diffuse galactic gamma-ray background originates primarily from the interaction of cosmic rays with the galactic matter and radiation fields.  
\cite{collaboration2020} construct a set of templates to fit galactic gamma-ray emission sources, and fit these to Fermi Large Area Telescope data as described in \cite{acero2015}\footnote{Updated information on the background modeling can be found at \url{https://fermi.gsfc.nasa.gov/ssc/data/analysis/software/aux/4fgl/Galactic_Diffuse_Emission_Model_for_the_4FGL_Catalog_Analysis.pdf}}.  
We refer to the resulting model as the non-isotropic background model.  
An isotropic background model is also fit to the data as a function of energy.  
The isotropic component includes unresolved extragalactic contributions, as well as the residual charged particle background.

We introduce the free parameter $b_{\rm iso}$ to scale the normalization of the isotropic background component.  
In effect, this parameter accounts for some uncertainty in the isotropic background model.
For our simulated data, the value of $b_{\rm iso}=1$ is used and we expect this value to be recovered in our analysis. 
For simplicity, we keep the normalization of the non-isotropic component fixed. 
\new{Of course, one clearly cannot adequately account 
for background mismodeling only by varying $b_{\rm iso}$.}
In \S\ref{sec:background_mismodeling}, we comment on the effects of mismodeling the galactic background. 
\new{There, we will consider only a rescaling of the 
amplitude of the anisotropic component, but background 
(mis)modeling may be much more complicated, and its 
effects would be an interesting topic of future work.}
We note that since the background models are informed by the observed gamma-ray sky, they could potentially already include a dark matter signal.  For the purposes of our analysis, however, this is not important as we are merely attempting to provide a rough characterization of the non-dark matter gamma-ray backgrounds.

\begin{figure}
    \centering
    \includegraphics[width=1.1\textwidth]{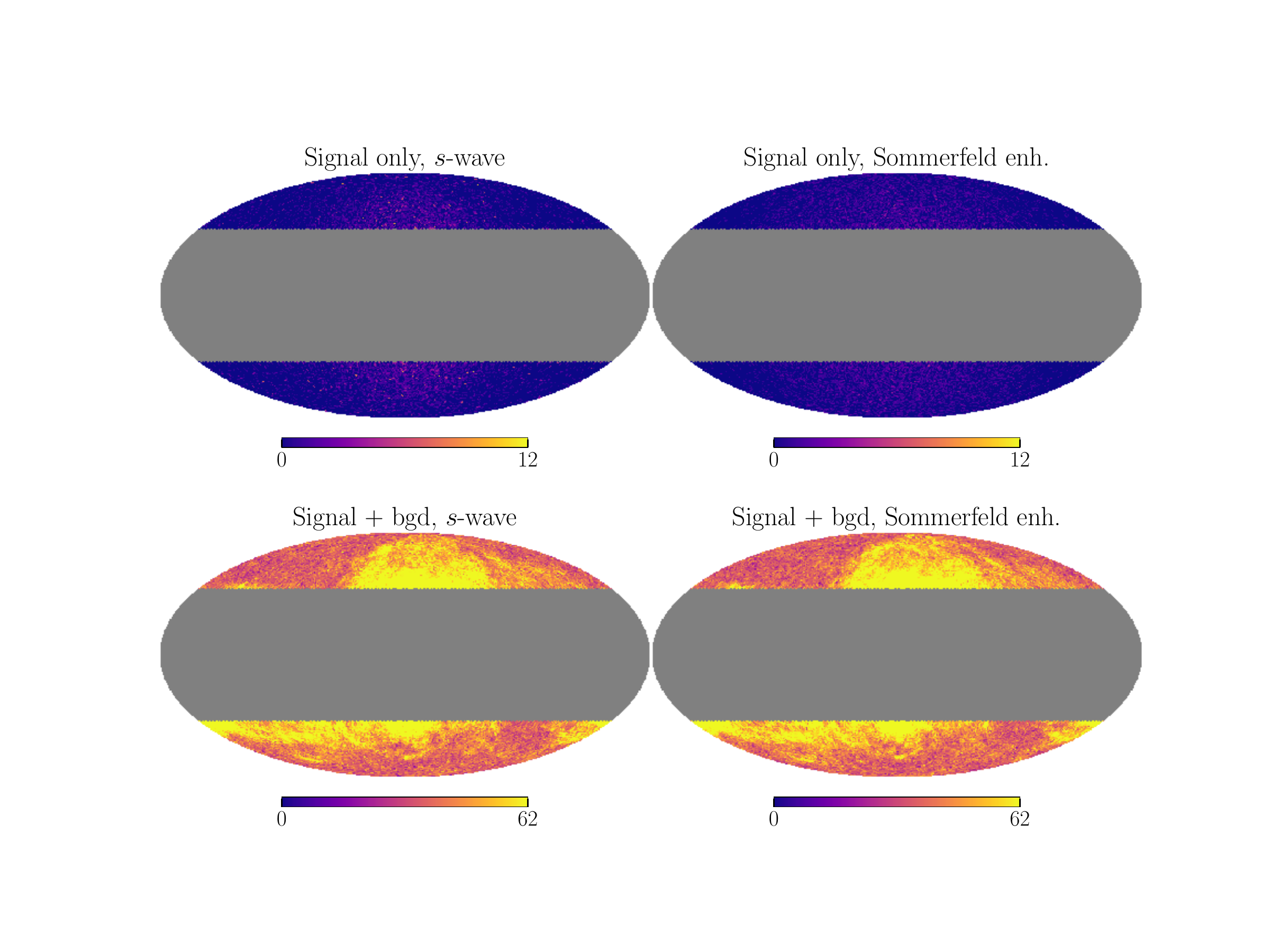}
    \caption{Simulated sky maps in Mollweide projection for $s$-wave (left) and Sommerfeld-enhanced (right) annihilation, with (top) and without (bottom) astrophysical backgrounds.  The Galactic Center is located at the center of the map, and horizontal corresponds to lines of constant galactic latitude.  The assumed exposure and dark matter parameters are detailed in \S\ref{sec:vel_dependence}, while the background model is described in \S\ref{sec:background_model}. \new{We take the normalization of the dark matter signal to be $\Phi_{PP}=  7 \times 10^{-30} \cm^3~\s^{-1}~\gev^{-2}$ consistent with the bounds obtained from Fermi data on dSphs \cite{Boddy:2019kuw}.}
    Consistent with Fig.~\ref{fig:pc}, $s$-wave annihilation leads to more frequent occurrence of bright pixels.  In our analysis, we mask a region within $40^{\circ}$ from the galactic plane (grey band) in order to minimize the impact of galactic backgrounds.      }
    \label{fig:skymaps}
\end{figure}

\section{Analysis of simulated data}
\label{sec:simulated_data}

We now generate and analyze mock data to illustrate how the statistics of the photon count distribution from dark matter annihilations in galactic substructure can be used to constrain the velocity dependence of the annihilation.  For simplicity, we assume that all the pixels are statistically independent.  In real data, this may not be the case as the instrument beam will correlate nearby pixels.  However, we do not expect this simplification to significantly change the conclusions of our analysis.  Moreover, the correlation between nearby pixels induced by the beam can always be reduced by increasing the pixel size.

We treat $n$ and $\Phi_{PP}$ as the main dark matter parameters of interest, although we will also explore the impact of varying $M_{\rm min}$ and $\beta$ in \S\ref{sec:degeneracy}.  Given the choice of these parameters described above, we compute $P_C(C)$ using Eq.~\ref{eq:p_of_c} for all the pixels of a \texttt{healpix} map with $N_{\rm side} = 128$, corresponding to a pixel size of $0.21\,\, \mathrm{deg}^2$. 
We first consider the signal-only case, for which we set $F^{bgd}_i=0$.  We draw from the resultant distributions to generate realizations of the dark matter signal, as seen in the top row of Fig.~\ref{fig:skymaps}.

Several features of the signal-only maps are readily apparent.  First, the signal increases in amplitude toward the galactic center, as expected from Eq.~\ref{eq:subhalo_luminosity}.  Second, one can clearly see the non-Poisson nature of the signal.  Rather than appearing as a smooth or uniform contribution to the sky, the signal-only maps are dominated by zero-count pixels, with occasional high-count pixels.  Finally, it is (somewhat) clear by eye that the $s$-wave map exhibits more non-Poisson flux than the Sommerfeld enhanced case, as expected from Fig.~\ref{fig:pc}.

Realizations that include backgrounds are generated in the same way, after setting $F^{bgd}_i$ to the flux from the background model discussed in \S\ref{sec:background_model}.
The combined signal and background map are shown in the bottom row of Fig.~\ref{fig:skymaps}.  
In each sky map, we mask the region with galactic latitude $|b|<40^\circ$ since this region will have very large galactic backgrounds. 
Note that the distinction between the $s$-wave and Sommerfeld maps that was visible by eye in the signal-only case is no longer visually 
clear when background is included, requiring a more 
detailed statistical analysis.

The mock data are analyzed as follows.  Given our assumption that the pixels are statistically independent, the likelihood for the data can be written as
\begin{equation}
    \mathcal{L}(\{C_i\}|\Phi_{PP}, b_{\rm iso}, n) = \prod_{i=1}^N P_C(C_i|\Phi_{PP}, b_{\rm iso}, n, i),
\end{equation}
where the product runs over all pixels in the map, $N$.  We will also consider an alternate model that removes information from the non-Poisson tail of $P_C(C)$.  We have
\begin{equation}
    \mathcal{L}^{\rm Poisson}(\{C_i\}|\Phi_{PP}, b_{\rm iso},n) = \prod_{i=1}^N \mathcal{P}(C_i | \bar{C}_{i}
    (\Phi_{PP}, b_{\rm iso},n)),
\end{equation}
where $\mathcal{P}(C_i|\bar{C}_{i})$ is the 
Poisson distribution with mean $\bar{C}_{i}$, where
\begin{equation}
    \bar{C}_{i}(\Phi_{PP}, b_{\rm iso},n) = 
    \sum_{j=0}^{\infty} 
    j P_C(j|\Phi_{PP}, b_{\rm iso}, n, i),
\end{equation}
is the expected number of photons in the $i$th pixel.
In other words, the Poisson model assumes that the distribution of photon counts is Poissonian with the same mean as the non-Poisson model.

More generally, we can express the expected number of photons 
per pixel as 
\bea
\langle N_{\rm true} (n) \rangle &=& \frac{1}{N} \sum_{i=0}^N \bar{C}_{i}(\Phi_{PP}, b_{\rm iso},n).
\eea
We will find it useful to exchange the parameter 
$\Phi_{PP}$ for $\langle N_{\rm true}(n) \rangle$.  
This parameter 
encodes the normalization of the dark matter 
annihilation signal (equivalently, the dark matter 
annihilation cross section), for any choice of $n$.

One of the goals of this analysis is to determine whether the velocity-dependence of the dark matter annihilation cross 
section can be determined from the statistics of the photon count distribution.  
To answer this question, we treat the determination of $n$ from the mock data as a model selection problem.  
Given two models for the velocity-dependence (specified by values of $n$), the likelihoods computed from these models can be maximized as a function of the model parameters.  
Then, the difference in maximum likelihoods for these two models --- specified by $n$ and $n'$ --- can be computed:
\begin{equation}
\Delta \ln \mathcal{L}_{\rm max} \equiv \max_{\Phi_{PP}, \,b_{\rm iso}} \left[ \mathcal{L}(\{ C_i\} | \Phi_{PP}, b_{\rm iso}, n) \right] - \max_{\Phi_{PP}, \,b_{\rm iso}} \left[ \mathcal{L}(\{ C_i\} | \Phi_{PP}, b_{\rm iso}, n') \right]. 
\end{equation}
\renew{Values of $\Delta \ln \mathcal{L}_{\rm max}$ can be used to compute model comparison statistics, such as the Akaike information criterion
and Bayesian information criterion (BIC) \cite{Elements}.  For the models considered here, we have that the difference in BIC for two models is $\Delta {\rm BIC} \sim -2 \Delta \ln \mathcal{L}_{\rm max}$.  The relative odds of the two models (under some approximations) is then $\exp[-\Delta {\rm BIC}/2] = \exp[-\Delta \ln \mathcal{L}_{\rm max}]$.  } Large values of $\Delta \ln \mathcal{L}_{\rm max}$ (that is, $ \gg 1$) would therefore indicate that the models can be distinguished at high significance from the data; small values, on the other hand, would suggest that there is not sufficient information in the data to distinguish between these possibilities.  In this analysis, we will consider two free parameters: 
the expected number of photons per pixel 
sourced by 
dark matter annihilation, $\langle N \rangle$,
and the normalization of the isotropic background component, $b_{\rm iso}$.
In an actual data analysis, uncertainty in astrophysical backgrounds is unlikely to be entirely captured by $b_{\rm iso}$.
However, this simple model allows for some uncertainty in the backgrounds, and serves to illustrate several important points.
We will consider the impact of additional \textit{systematic} uncertainty in the background model in \S\ref{sec:background_mismodeling}.

\section{Results}
\label{sec:results}

\begin{figure}
    \centering
    \includegraphics[width=\textwidth]{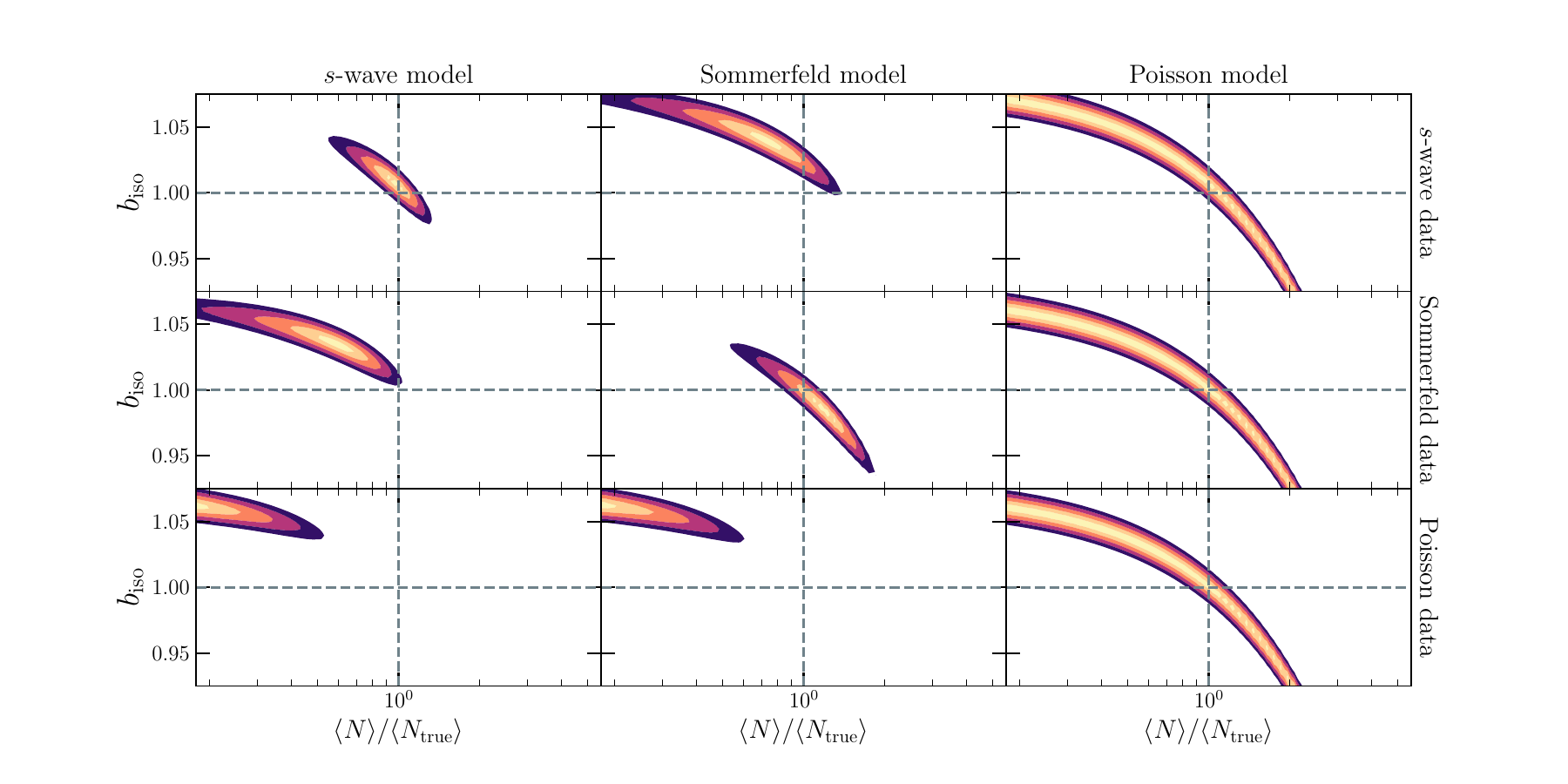}
    \caption{Parameter constraints from the mock data analysis described in \S\ref{sec:simulated_data}.  The three rows correspond to data generated using three different models: $s$-wave (top), Sommerfeld enhancement (middle), and Poisson (bottom). The contours correspond to 1-, 2-, 3-, 4-, and 5-sigma. In all three cases, we choose the normalization of the dark matter signal to yield the same mean (see \S\ref{sec:simulated_data}).
    The different columns correspond to analyzing the data assuming different models (see labels at top).  We vary both the normalization of the dark matter signal, $\langle N \rangle/\langle N_{\rm true} \rangle$ (i.e. the mean number of photons produced by dark matter annihilation relative to the true mean), and the normalization of the isotropic background, $b_{\rm iso}$.  We see that when the data are analyzed using the correct model, we recover the true parameter values to within the errors (panels along diagonal).  We also see some degeneracy between the dark matter signal normalization and the amplitude of the isotropic background.  For the Poisson model, this degeneracy is more severe since in this case, the dark matter signal and the isotropic background have the same (Poisson) photon count distribution.  }
    \label{fig:contours}
\end{figure}

\begin{table}[]
    \centering
    \begin{tabular}{|c||c|c|c|}
    \hline
         True model & v.s. free $b_{\rm iso}$ 
         + Poisson& v.s. free $b_{\rm iso}$ + $s$-wave & v.s. free $b_{\rm iso}$ +  Som. \\
    \hline
        Poisson & ---  & 35.9 & 19.9 \\
    \hline
         $s$-wave & 21.3 & --- & 35.5 \\
    \hline
         Sommerfeld & 24.7 & 46.3 &  --- \\
    \hline
    \end{tabular}
    \caption{The values of $\Delta \ln \mathcal{L}_{\rm max}$ associated with the indicated model comparisons, computed from mock data.  The left column indicates the true model, while the top row indicates the alternate model.  Values along the diagonal are 0 by definition.  We show values assuming mock data with the current Fermi exposure and with $\langle N_{\rm true} \rangle = 0.37$. 
    We see that the correct model is preferred with high significance.}
    \label{tab:deltaloglike}
\end{table}

\subsection{Ability to distinguish different velocity-dependence models}
\label{subsec:ability}

In Fig.~\ref{fig:contours} we show constraints on the two parameters of our model --- 
$\langle N \rangle$
and $b_{\rm iso}$ --- from the analysis of the mock data.  
We consider three sets of mock data, in 
which the data are generated assuming either the $s$-wave, 
Sommerfeld, or Poisson models.  \renew{Each set of mock data corresponds to a single realization of the sky signal, as would be the case for real data.  } For all three cases, the 
luminosity normalization is chosen so that the average 
expected number of photons per pixel is fixed to 
$\langle N_{\rm true} \rangle =0.37$, which 
corresponds to a $\Phi_{PP}$ normalization roughly at the 
limit of Fermi searches for gamma rays from dark matter 
annihilation in dSphs. 

Note that
$\bar C (\psi_i ; n) = E_i \mu (\psi_i) \int dF ~ 
F \times P_1 (F ; \psi_i ; n)$, where the dependence of 
$\bar C$ on $n$ essentially factorizes from its dependence 
on $\psi_i$, because the dependence on $n$ arises only 
from the luminosity distribution $P_L (\ln L_{sh})$, which 
is nearly independent of $\psi_i$.  Thus, normalizing the 
$s$-wave and Sommerfeld models such that they produce the same number 
of photons averaged over all pixels means that the expected photon counts in each pixel will also be the same for the two models, i.e. $\bar C (\psi_i ; n=0) = \bar C (\psi_i ; n=-1)$ 
for every pixel $i$.  However, even if the expected photon counts for two models are matched, the photon count {\it distributions} for each pixel will not be the same in the two models.
This scenario is therefore very different from studies 
of the velocity-dependence of dark matter annihilation 
in dSphs~\citep{Baxter:2021pzi}.  In that case, since one is observing resolved 
dark matter subhalos with particular halo parameters, 
a change in the velocity dependence of dark matter 
annihilation can lead to changes in  the expected photon 
counts from various dSphs which cannot be compensated by 
an overall normalization change.
Information about the velocity dependence of the dark matter annihilation is therefore contained in the relative fluxes from dSphs.  In the present case, differences in the expected photon counts for two models can be compensated by an overall normalization change, so information about the velocity dependence of the dark matter annihilation must be extracted from the (non-Poisson) photon count distribution in each pixel.

We see from Fig.~\ref{fig:contours} that when the data are analyzed with the correct model (i.e. the panels along the diagonal), the input parameters are recovered to within the errorbars.  The precision of the constraints is also encouraging: even with current Fermi data, there appears to be sufficient statistical power in the data to constrain the dark matter signal normalization to high precision.  This is consistent with the findings of \cite{Baxter:2010fr}.  \new{We note that this does not mean current Fermi data can be used to constrain the normalization to high precision -- making such a statement would require more sophisticated background modeling than we are attempting here \cite[e.g.][]{Somalwar:2020awt}.  Rather, this means that there is \textit{in principle} sufficient information in the data, and further work to develop background models for an actual data analysis is motivated.}

We see that there is significant anti-correlation between the signal amplitude and the isotropic background normalization parameter, $b_{\rm iso}$.  This is not too surprising: more signal photons can be partially compensated by a reduced isotropic background.  In the case of Poisson model analysis, this degeneracy is extreme because both the isotropic background model and the Poisson dark matter model are described by Poisson distributions.\footnote{Note that the degeneracy must be broken at some level, since the Poisson dark matter model decreases in amplitude away from the Galactic Center, while the isotropic background is uniform across the sky.}  The fact that this degeneracy is less severe when the data are analyzed with the $s$-wave and Sommerfeld models suggests that the non-Poisson information in these models is contributing significantly to the constraints.

When the data are analyzed with the incorrect dark matter model (off-diagonal panels of Fig.~\ref{fig:contours}), the recovered parameters constraints are generally biased.  
In this case, we do not recover the input parameters, even when the dark matter model is correct.

The fact that this bias is small when the data are analyzed with the Poisson model likely reflects the fact that this model effectively discards non-Poisson information.
With only the Poisson information, all that matters is the total photon counts, and these will only be recovered when $\langle N \rangle  = \langle N_{\rm true} \rangle$.
The bias becomes severe in the non-Poisson cases.

In Table~\ref{tab:deltaloglike} we present the $\Delta \ln \mathcal{L}_{\rm max}$ values for the model comparisons computed with the mock data.  Values of $\Delta \ln \mathcal{L}_{\rm max} \gg 1$ indicate that
in each case, even the current Fermi 
exposure is sufficient to distinguish the true velocity-dependence model from an incorrect 
model at high significance.  As expected, it is most 
difficult to distinguish between Sommerfeld-enhanced 
annihilation and a Poisson-distributed photon source, because the photon count distribution arising from Sommerfeld-enhanced 
annihilation in subhalos looks close to Poisson.  \renew{We note that since the $\Delta \ln \mathcal{L}_{\rm max}$ are computed from a single stochastic realization of the data, they are inherently noisy.  However, the expected variance in $\Delta \ln \mathcal{L}$ is $\sim 1$, significantly smaller than the reported values.}

\subsection{Sensitivity to background modeling assumptions}
\label{sec:background_mismodeling}

\begin{figure}
    \centering
    \includegraphics[width=\textwidth]{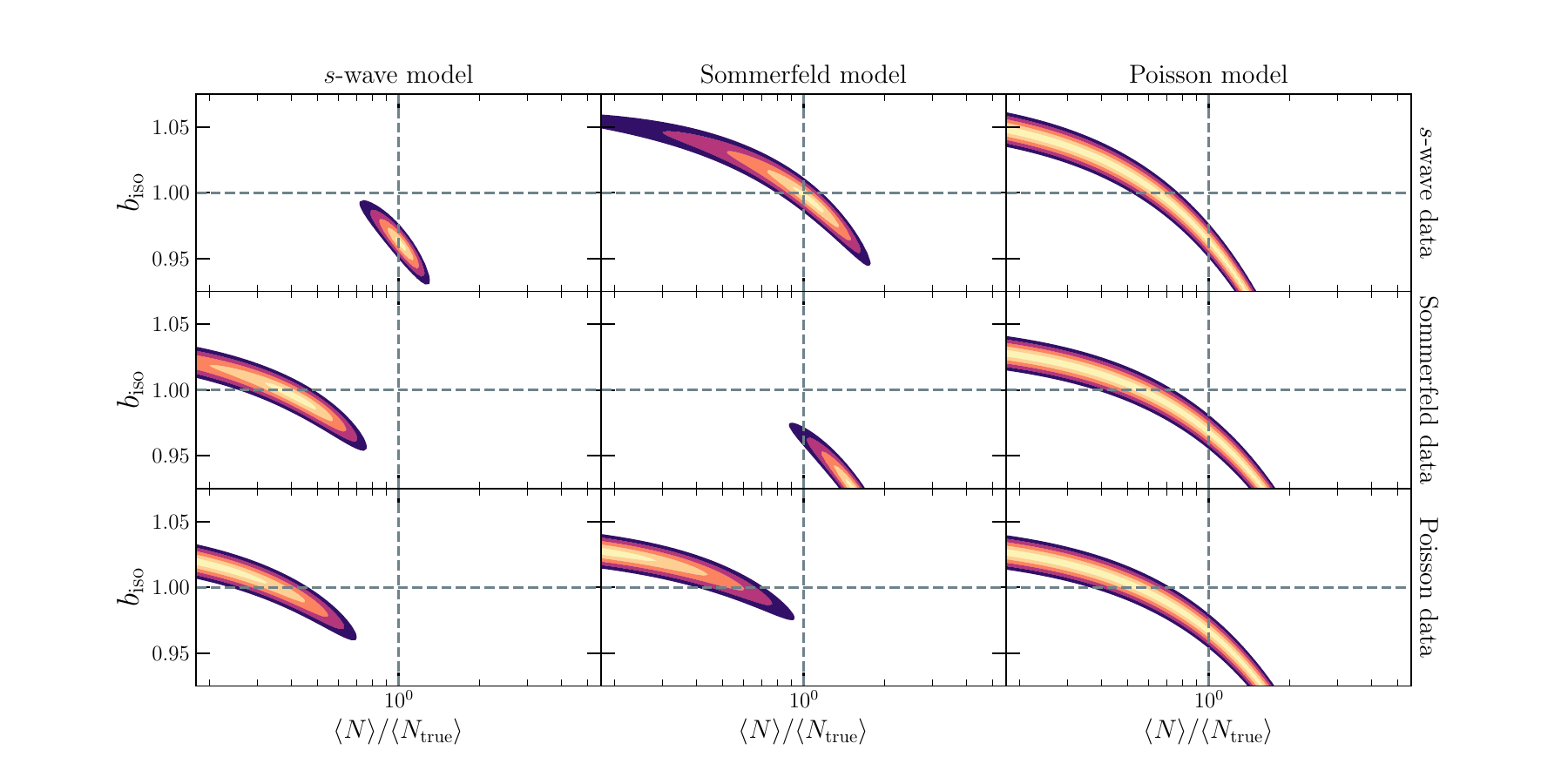}
    \caption{Same as Fig.~\ref{fig:contours}, except the model for the non-isotropic galactic backgrounds used to compute likelihoods has a 
    normalization that is $3\%$ larger than the 
    normalization 
    of the mock data 
    (see \S\ref{sec:background_mismodeling}). 
    }
    \label{fig:contours_mismodelled}
\end{figure}

\begin{table}[]
    \centering
    \begin{tabular}{|c||c|c|c|}
    \hline
         True model & v.s. free $b_{\rm iso}$ 
         + Poisson & v.s. free $b_{\rm iso}$ + $s$-wave & v.s. free $b_{\rm iso}$ +  Som. \\
    \hline
        Poisson & ---  & 35.3 & 19.2 \\
    \hline
         $s$-wave & 43.5 & --- & 55.3 \\
    \hline
         Sommerfeld & 49.9 & 67.4 &  --- \\
    \hline
    \end{tabular}
    \caption{The values of $\Delta \ln \mathcal{L}_{\rm max}$ associated with the indicated model comparisons, where the model has a a mismodelled galactic background at 103\% of the `true' value.  The left column indicates the true model, while the top row indicates the alternate model.  Values along the diagonal are 0 by definition.  We show values assuming mock data with the current Fermi exposure and with $\langle N_{\rm true} \rangle = 0.37$.  
    We see that the correct model is preferred with high significance.}
    \label{tab:deltaloglike_mismodelled}
\end{table}

Thus far, we have assumed that uncertainty in our model of astrophysical backgrounds is fully encapsulated in the scaling of the isotropic background component through the parameter $b_{\rm iso}$.  
Of course, this is unlikely to be the case in any real data analysis, given the complexity of the galactic gamma-ray backgrounds.  
\renew{We now consider several specific examples of what can happen when the non-isotropic
backgrounds are mismodelled.  In these examples, we consider mock data generated 
as described in \S\ref{subsec:ability}, 
using the 
background model described in \S\ref{sec:background_model}.  But we will evaluate the likelihoods assuming different (i.e. incorrect) models for this background.  First, we adopt a model for which the} amplitude of the anisotropic 
background is 3\% larger than that of the true background.  In the second case, 
likelihoods are evaluated assuming 
that the amplitude of the anisotropic 
background is 3\% smaller.  \renew{Finally, we consider a case where the assumed background model has a different spatial resolution than the true background.
These simple examples are meant to explore our sensitivity to possible systematic errors in the modeling of the gamma-ray backgrounds.  }

In the case where the anisotropic background model 
overestimates the true anisotropic background, 
the resultant parameter constraints are shown in Fig.~\ref{fig:contours_mismodelled}.
In Table~\ref{tab:deltaloglike_mismodelled} we present the associated $\Delta \ln \mathcal{L}_{\rm max}$ 
values.  We find that if mock data is generated with $s$-wave annihilation, then an analysis 
assuming $s$-wave annihilation can recover the correct normalization of the dark matter signal 
even when the anisotropic background is mismodeled.
\renew{The recovered normalizations for the isotropic background do not reflect the `true' value, rather they need to compensate for the anisotropic background mismodeling.
This inaccuracy can be anticipated because the likelihood will be maximized for parameter combinations that yield the same number of photons as the data. 
In this case, the model yields an expected number of photons from non-isotropic galactic backgrounds which is larger than in the mock data, so likelihood is maximized for parameter choices which yield fewer photons from isotropic backgrounds.}
In this case, the effect of the overestimated anisotropic background is to suppress the recovered normalization of the isotropic background.
On the other hand, if mock data is generated with Sommerfeld-enhanced annihilation, an analysis assuming Sommerfeld-enhanced annihilation and a mismodeled anisotropic background would recover a dark matter signal which is too large, and an isotropic background whose normalization is too 
small.\footnote{Note, though, that if mock data is 
generated assuming $s$-wave dark matter annihilation, but analyzed assuming Sommerfeld-enhanced 
annihilation, the correct normalization of the dark matter signal is recovered.  This appears to 
be a coincidence.}

But, perhaps unexpectedly, we find that, even if the anisotropic 
background is overestimated, 
it is still possible to distinguish the correct velocity-dependence of dark 
matter annihilation.  In particular, if the mock data are generated assuming a  
Poisson-distributed signal, then that model will be preferred over either $s$-wave or Sommerfeld-enhanced 
annihilation, even with a mismodeled background.  This may seem counterintuitive, and contrary to 
some of the lessons learned from studies of the GC excess, because a Poisson fluctuation about the 
true mean may appear to be a non-Poissonian fluctuation about a mismodeled mean.  In this case, model discrimination is possible because we 
are not looking for non-Poisson fluctuations, but rather are 
comparing the data to a particular non-Poisson 
distribution.

\begin{figure}
    \centering
    \includegraphics[width=\textwidth]{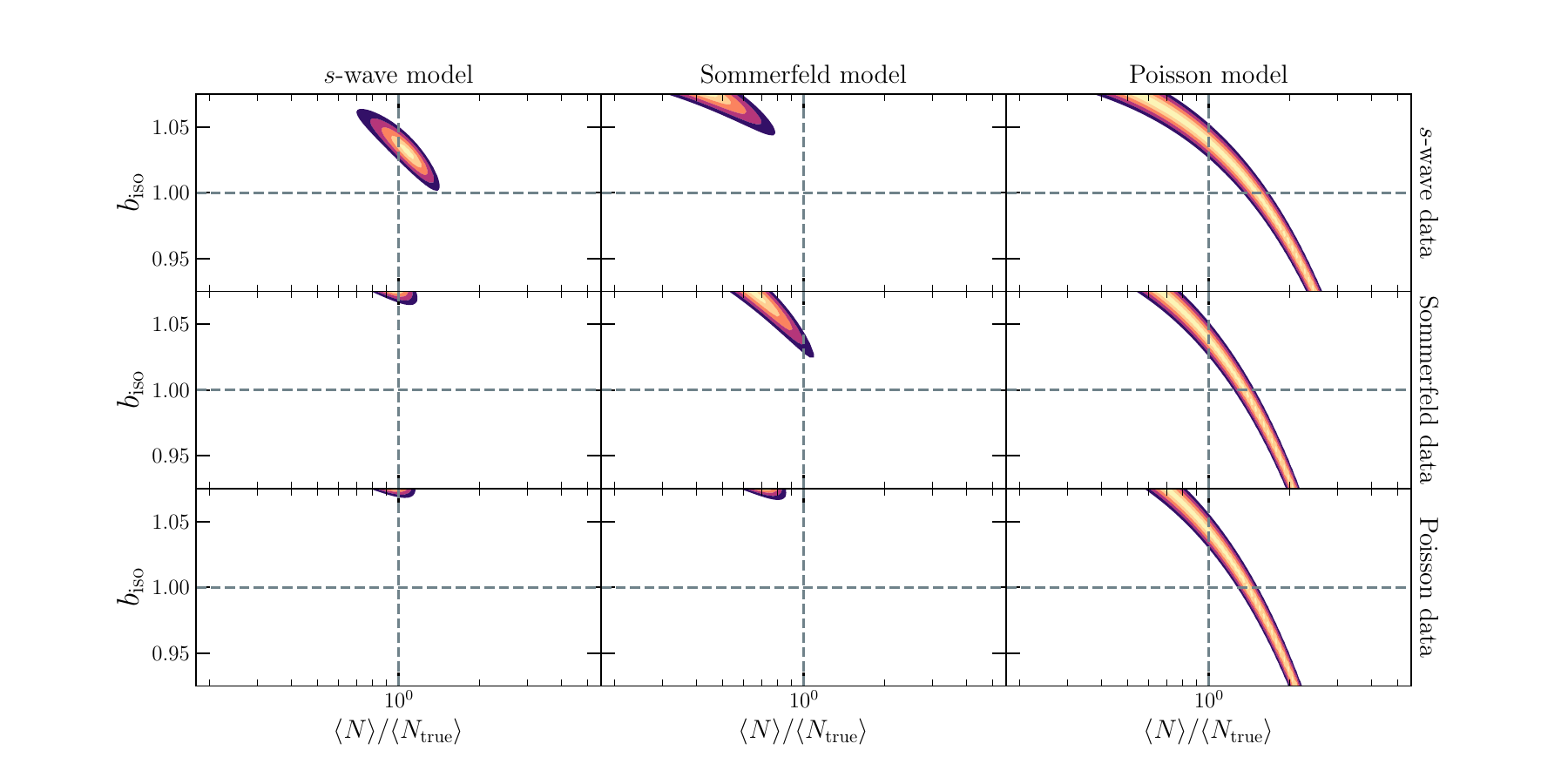}
    \caption{Same as Fig.~\ref{fig:contours_mismodelled}, but for the case when the model for the non-isotropic galactic backgrounds is  $3\%$ smaller than the normalization 
    of the mock data 
    (see \S\ref{sec:background_mismodeling}).  As in 
    Figure~\ref{fig:contours_mismodelled}, we do not recover the input parameters, even when the dark matter model is correct.  
    }
    \label{fig:contours_mismodelled_v2}
\end{figure}

\begin{table}[]
    \centering
    \begin{tabular}{|c||c|c|c|}
    \hline
         True model & v.s. free $b_{\rm iso}$ 
         + Poisson & v.s. free $b_{\rm iso}$ + $s$-wave & v.s. free $b_{\rm iso}$ +  Som. \\
    \hline
        Poisson & ---  & 36.1 & 20.1 \\
    \hline
         $s$-wave & 1.3 & --- & 17.6 \\
    \hline
         Sommerfeld & 5.2 & 29.9 &  --- \\
    \hline
    \end{tabular}
    \caption{The values of $\Delta \ln \mathcal{L}_{\rm max}$ associated with the indicated model comparisons, fit with a model with a mismodelled galactic background at 97\% of the value used to generate the data.  
    The left column indicates the true model, while the top row indicates the alternate model.  Values along the diagonal are 0 by definition.  We show values assuming mock data with the current Fermi exposure and with $\langle N_{\rm true} \rangle = 0.37$.  
    We see that the correct model is again preferred, but 
    with lower significance than in the case of an 
    overestimated background. 
    }
    \label{tab:deltaloglike_mismodelled_undermodelled}
\end{table}

{For the case in which the anisotropic 
background model underestimates the true anisotropic 
background, we plot the parameter constraints in 
Figure~\ref{fig:contours_mismodelled_v2}, and we 
list the values of $\Delta \ln \mathcal{L}_{\rm max}$ in 
Table~\ref{tab:deltaloglike_mismodelled_undermodelled}.  
These results provide an interesting contrast to the 
case in which the background is overestimated.  
In either case, one is able to 
distinguish the true model for the 
dark matter signal from an incorrect non-Poisson model, 
though the significance is weaker when the background 
model underestimates the true background.  
But if the background model underestimates the true background, then it is difficult to reject 
a Poisson model for the dark matter signal; 
even if the true model is non-Poisson, the Poisson model 
will be only slightly disfavored.  The 
reason is because, if the anisotropic background model 
underestimates the true anisotropic background, then 
there is a significant excess of photons, beyond that 
predicted by the background model, which is 
Poisson-distributed but anisotropic.  
This excess is in addition to any 
photons arising from dark matter annihilation.  Although 
the best fit arises from increasing the amplitude 
of the Poisson-distributed isotropic background and 
adding a source with the correct non-Poissonian 
distribution, 
the log likelihood is only slightly smaller if one 
adds a Poisson-distributed 
source of photons with the same spatial distribution as 
the dark matter signal.
}

In Table~\ref{tab:deltaloglike_dif_bgds}, we compare the maximum likelihood obtained 
for each true model of dark matter annihilation, when 
analyzed with the correct dark matter model and correct 
background model, as compared to the correct dark matter 
model but incorrect background model, with the anisotropic 
background either underestimated or overestimated by 
$3\%$.  As expected, we find a preference for the model with 
the correct anisotropic background normalization.  But 
interestingly, in the case where the dark matter signal 
photons are Poisson-distributed, the model with a 
correctly-modelled anisotropic background is only slightly 
preferred to the model in which the anisotropic background 
is overestimated.

\begin{table}[]
    \centering
    \begin{tabular}{|c||c|c|}
    \hline
        Model & underestimated & 
        overestimated\\
    \hline
        Poisson & 51.6 & 2.4 \\
    \hline
         $s$-wave & 21.0 & 30.9 \\
    \hline
         Sommerfeld & 35.5 &  13.5 \\
    \hline
    \end{tabular}
    \caption{Values of $\Delta\ln\mathcal{L}_{\rm max}$ when the true model with a correctly modelled background is compared to the same dark matter model, but with the anisotropic 
    background underestimated (first column) or overestimated (second column) by $3\%$.
    }
    \label{tab:deltaloglike_dif_bgds}
\end{table}

\renew{We now consider the case where the adopted background model has a lower spatial resolution than the backgrounds used to generate the data.  A background model can never hope to perfectly capture the full complexity and small-scale structure of the true diffuse gamma-ray sky.  The scenario we explore here is meant to test the impact of removing some information in the background model through coarse-graining.  Previous analyses \citep{Leane:2019xiy,Leane:2020nmi} have found that similar smoothing of adopted background templates can increase the significance of non-Poisson source detections. 
To generate the smoothed background model, we decrease the resolution of the fiducial \texttt{healpix} skymap by degrading the resolution from $N^{\rm fiducial}_{\rm side}=128$ to $N_{\rm side}^{\rm smoothed}$ with $N_{\rm side}^{\rm smoothed} < N^{\rm fiducial}$, and then resampling the resultant map back to the original resolution, thereby averaging the background template pixels over the neighboring pixels. 
The effects of this smoothing on the $\Delta \ln\mathcal{L}_{\rm max}$ values is shown in Table~\ref{tab:deltaloglike_smoothed_bgd}.
We see a similar result to \cite{Leane:2020nmi}: as we smooth the anisotropic background model, we see increasing preference for the $s$-wave model (which is highly non-Poisson) when the true model is $s$-wave.  In other words, when there is a non-Poisson signal in the data, using a smoothed background model can enhance the preference for that non-Poisson signal.
Interestingly, we do not see a similar effect with the Sommerfeld-enhanced data nor with the Poisson data.
The $\Delta\ln\mathcal{L}_{\rm max}$ values increase with increasing smoothness for $s$-wave data but remain nearly constant for Sommerfeld and Poisson data.  In other words,  if there is only a weak non-Poisson signal in the data, then using a smoothed background model does not significantly change the preference for non-Poisson signals.
}

\renew{The different perturbations to the background model that we consider above explore some of the effects of background mismodelling on our analysis.  Given the large number of degrees of freedom in the background model, there are effectively an infinite number of ways that the backgrounds could conceivably be mismodeled.  To ensure robustness of results in an analysis on data, one could perform additional tests, such as repeating the analysis on different patches of the sky with different levels of expected background, and comparing the results. }

\begin{table}[]
    \centering
    \begin{tabular}{|c||c|c|c|}
        \cline{0-0}
        True Model \\
        \hline
        $N_{\rm side}^{\rm fiducial}=128$ & v.s. free $b_{\rm iso}$ + Poisson & v.s. free $b_{\rm iso}$ + $s$-wave & v.s. free $b_{\rm iso}$ +  Som. \\
    \hline
        Poisson & ---  & 35.94 & 19.95 \\
    \hline
        $s$-wave & 19.7 & --- & 26.9 \\
    \hline
        Sommerfeld & 30.7 & 56.5 &  --- \\
    \hline
    \hline
        $N_{\rm side}^{\rm new}=64$ &&& \\
    \hline
        Poisson & --- & 35.90 & 19.91 \\
    \hline
        $s$-wave & 27.9 & --- & 37.5 \\
    \hline
        Sommerfeld & 31.2 & 55.7 & --- \\
    \hline
    \hline
        $N_{\rm side}^{\rm new}=32$ &&& \\
    \hline
        Poisson & --- & 35.88 & 19.95 \\
    \hline
        $s$-wave & 45.6 & --- & 53.7\\
    \hline
        Sommerfeld & 29.6 & 51.3 & --- \\
    \hline
    \end{tabular}
    \caption{\renew{Values of $\Delta\ln\mathcal{L}_{\rm max}$ when the true model with a correctly modelled background is compared to the same dark matter model, but with a coarse-grained version of the anisotropic 
    background model (with resolution indicated by the value of $N_{\rm side}$ in the table).}}
    \label{tab:deltaloglike_smoothed_bgd}
\end{table}

\subsection{Additional parameter degeneracies}
\label{sec:degeneracy}

\begin{figure}
    \centering
    \includegraphics[width=0.6\textwidth]{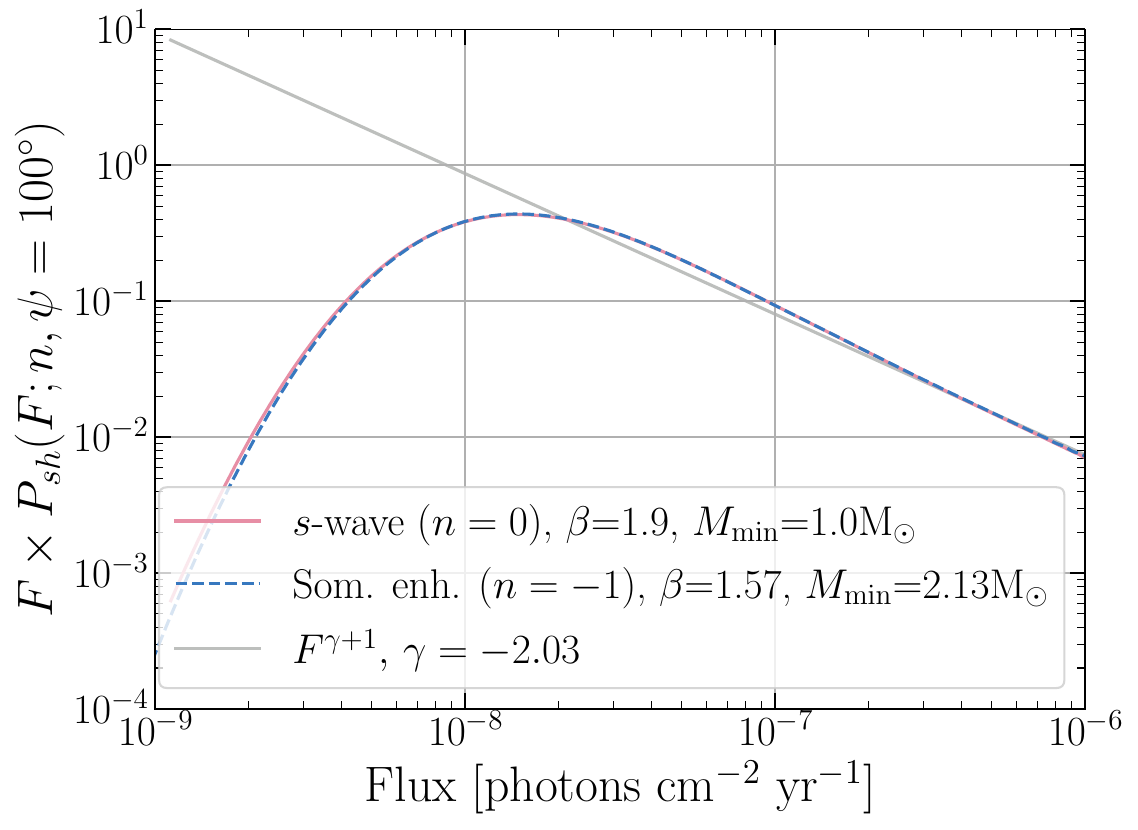}
    \caption{
    Illustration of the degeneracy between $\beta$, $M_{\rm min}$, and $n$ discussed in \S\ref{sec:degeneracy}.  We show curves of $P_{\rm sh}(F ; n)$ ($n=-1, 0$ are shown with the blue dashed and red solid lines, respectively).  For $n=-1,0$, we choose $\beta = 1.57, 1.9$, respectively, so as to yield $\gamma = -2.03$ ($\gamma$ controls the high-$F$ slope of $P_1(F)$ and $P_{\rm sh}(F)$, as indicated by the solid grey line).  
    For $n=-1, 0$ we choose $M_{\rm min} = 2.13M_{\odot}, 1.0M_\odot$, respectively,  so that in both 
    cases we have $\mu(\psi_i=100^\circ)\approx 51800$.  With these choices, we find that $P_{\rm sh}(F ;n=0)$ 
    and $P_{\rm sh}(F ;n=-1)$ are nearly identical. 
    }
    \label{fig:degeneracy_illustration}
\end{figure}

Our analysis above has only varied the dark matter parameters
$\langle N \rangle$
and $n$ when fitting the mock data.  We now explore parameter degeneracies that can impact our ability to uniquely determine $n$ from measurement of the photon count distribution.  

The dark matter signal flux distribution, $P_{sh}(F)$, is determined by $P_1(F)$ (the flux probability distribution for a single subhalo) and $\mu$ (the expected number of subhalos in a pixel) via Eq.~\ref{eq:p1_to_pf}.  The high flux tail of $P_1(F)$ is in turn determined by $\gamma$ (Eq.~\ref{eq:gamma}), which  depends on the subhalo mass function (Eq.~\ref{eq:mass_func}) and 
luminosity (Eq.~\ref{eq:param_dep}).  It is this high flux tail which sets the non-Poisson tail of the photon counts distribution.

Consider two models for the dark matter annihilation velocity dependence characterized by different values of $n$.  If we adjust model parameters so that $\gamma$ is the same for these models, then the two models will predict the same power law behavior of $P_{sh}(F)$ at high $F$.  Obtaining identical values of $\gamma$ for different values of $n$ can be accomplished by adjusting the $\beta$ or $\alpha$ values for the two models.  If we also adjust parameters (such as $M_{\rm min}$) so that $\mu$ is the same for both models, then we would expect the resultant $P_{sh}(F)$ to be very similar, up to an overall normalization.  Since we would typically treat the overall flux normalization as a free parameter (since it depends on the unknown properties of the dark matter), it would then be very difficult to distinguish these models from the data.

We explore this possibility in Fig.~\ref{fig:degeneracy_illustration}.  Here we have adjusted $\beta$ to keep $\gamma = -2.03$, and have adjusted $M_{\rm min}$ to keep $\mu(\psi_i=100^\circ) \approx 51800$.  For the curves 
with $n=-1$ and $n = 0$, we use $\beta = 1.57$ and $\beta = 1.9$ with
$M_{\rm min} = 2.13\,M_\odot$ and $M_{\rm min} 1.0\,M_\odot$, respectively, while 
maintaining $M_{\rm max} = 10^{10}\,\,\mathrm{M_\odot}$ in 
both cases.  
Once these adjustment are made, we scale the
luminosity normalization to match 
the mean flux of the two distributions.
The agreement between the three curves in Fig.~\ref{fig:degeneracy_illustration} confirms the expected degeneracy between $M_{\rm min}$, $\beta$, $n$, and the subhalo luminosity 
normalization (equivalently, the normalization of the 
dark matter annihilation cross section).  
We note that while Fig.~\ref{fig:degeneracy_illustration} varies $\beta$ as an example, a similar result would be achieved by varying $\alpha$, as both parameters impact $\gamma$.  
Moreover, similar degeneracies could result if additional freedom were introduced into the subhalo mass function and mass-luminosity relation.  
In general, since changing $n$ changes the high-flux tail of $P_1(F)$, any variation of parameters that also modifies the high flux tail of $P_1(F)$ while preserving $\mu$ could result in a degeneracy that interferes with our ability to measure $n$.
\new{In the case of degenerate PDFs, neither model will be preferred over the other because the PDF is the only information entering our likelihood analysis.
Thus, when the PDFs are degenerate, our analysis is unable to determine the velocity dependence of dark matter annihilation.}

The impact of these parameter degeneracies on our ability to determine the velocity dependence of the dark matter annihilation can be ameliorated with prior information on the mass function or mass-luminosity relation.  Changing $n$ results in a significant change to $P_{sh}(F)$, and counteracting these changes requires fairly extreme variations in other parameters.  For example, we needed to change $\beta$ from $1.9$ to $1.57$ to produce the overlapping $P_{sh}(F)$ shown in Fig.~\ref{fig:degeneracy_illustration}.  Consequently, with tight priors on parameters such as $\beta$, the impact of the parameter degeneracies highlighted above will be reduced.  Prior information on the subhalo mass function and mass luminosity relation can be obtained, for instance, from high resolution $N$-body simulations \cite[e.g.][]{Fiacconi:2016}.  Observational constraints from, for instance, strong gravitational lensing \citep{Xu:2009,Vegetti:2010,Ostdiek:2020}, can also provide prior information on the subhalo mass function.

\section{Conclusion}
\label{sec:conclusion}

We have shown that the statistics of photons produced in dark matter annihilation in galactic subhalos carries information about the velocity dependence of the annihilation of the dark matter. 
As seen in, e.g., Fig.~\ref{fig:p1psh}, $s$-wave annihilation ($n=0$) in subhalos leads to a flatter tail in the probability distribution for the flux along a given line of sight than Sommerfeld enhanced annihilation ($n=-1$).  
For larger $n$ models --- e.g., $p$-wave ($n=2$) and $d$-wave ($n=4$) annihilation --- the tail will be even more pronounced.  
As a result, models with large $n$ will result in \new{the counts distribution having a much stronger tail to high count values than a Poisson distribution}.
The fundamental reason for this can be seen in Eq.~\ref{eq:param_dep}: larger $n$ causes the subhalo luminosity to increase faster with subhalo mass, leading to a higher probability of each line of sight having a bright subhalo.

Our results suggest that in principle, the photon counts distribution can be used to determine the velocity dependence of dark matter annihilation.
Indeed, Table~\ref{tab:deltaloglike} show that there is sufficient statistical power in current data to rule out particular velocity-dependence models.  

However, we have also identified several challenges to this program.
First, there are degeneracies between the velocity dependence of the annihilation cross section, the subhalo mass function, and the subhalo mass-luminosity relation.
These three functions all impact the high flux tail of the subhalo flux probability distribution function.
As seen in Fig.~\ref{fig:degeneracy_illustration}, with appropriate choices of $M_{\rm min}$ and $\Phi_{PP}$, a near perfect degeneracy results.
Consequently, in order to definitively identify the velocity dependence of the dark matter annihilation, we must have robust predictions for the subhalo mass function and mass-luminosity relation.
Such predictions can be obtained from $N$-body simulations and potentially observations of e.g. gravitational lensing.

A second, perhaps more severe, challenge is that using the photon counts distribution to determine the velocity dependence of the dark matter annihilation will require a precise understanding of astrophysical backgrounds.  In contrast to diffuse galactic backgrounds, subhalos generically cause non-Poisson behavior in the photon counts probability distribution function \citep{Lee:2008fm, Baxter:2010fr}.  However, this difference in the count distributions is not necessarily sufficient to distinguish signal from backgrounds, given the much larger amplitude of the backgrounds.  
Previous analyses of the GC excess have 
shown that, even if the photon source is Poisson-distributed, mismodeled backgrounds can be misinterpreted as a 
non-Poisson-distributed source.  But we have shown that, 
when one compares the likelihoods of a Poisson-distributed 
source model to a particular non-Poisson-distributed source 
model, the ability to distinguish the true model is at least 
somewhat robust against background mismodeling (see 
Tables~\ref{tab:deltaloglike_mismodelled} and~\ref{tab:deltaloglike_mismodelled_undermodelled}).

In particular, even if the true model 
includes a non-Poissonian photon source, it seems 
possible to 
reject a model with a different 
non-Poisson distribution.  On the other hand, if 
the background model leaves a Poisson-distributed 
background photon source unaccounted for, then 
it may be difficult to reject a Poisson model 
for the photons arising from dark matter annihilation, 
even if the photons sourced by dark matter annihilation 
are actually non-Poissonian.  
\new{However, even if it is 
possible to correctly determine velocity-dependence 
in the presence of a mismodelled background, it is 
would generally be much more difficult to determine 
the normalization of the dark matter signal.}

But note that we performed a
relatively simple analysis, assuming only a one-parameter 
background photon distribution. 
The only allowed variation in the normalization of 
the Poisson-distributed background is spatially 
isotropic.  
\new{Moreover, it is clear that mismodeling of background cannot be fully accounted for by only varying the 
amplitude of the anisotropic component.}
It would be interesting to see if, with a more 
complete background model, one could better distinguish the 
presence of a non-Poisson distributed dark matter source 
from the effects of a mismodelled anisotropic Poisson background.  \new{It would also be interesting to 
study the robustness of these results in the presence 
of more realistic versions of background mismodeling.}

Furthermore, in this analysis, we have ignored astrophysical backgrounds from unresolved point sources --- such as star forming galaxies and blazars --- 
which are likely to also interfere with our ability to identify the dark matter signal.  
But the fact that we have demonstrated 
an ability to distinguish between two different 
non-Poisson-distributed source models, even in the 
presence of some background mismodeling, gives hope that 
one can also discriminate the correct model for dark matter 
annihilation velocity dependence even in the presence of 
non-Poisson-distributed astrophysical backgrounds, 
\new{provided the amplitude of the dark matter signal is 
large enough.  A similar question was considered in~\cite{Somalwar:2020awt}, which found that 
one could detect the presence of a non-Poisson 
photon signal arising from $s$-wave 
dark matter annihilation in 
subhalos in the presence of a non-Poisson signal arising 
from blazars, given a sufficiently large dark matter 
signal.  
It would be interesting to perform a more comprehensive analysis, including a realistic 
distribution of unresolved astrophysical point sources, 
along with realistic uncertainties in both Poisson 
and non-Poisson astrophysical backgrounds.}

Finally, we note that our results indicate 
that, with current Fermi data, \new{there is sufficient statistical information} to detect evidence 
for dark matter annihilation in subhalos with a 
cross section which is roughly at the sensitivity of 
dSph searches.  
\new{
These seems somewhat more optimistic than the results 
found in Ref.~\cite{Somalwar:2020awt}, which suggested 
that searches of Fermi data from dSphs significantly 
exceeded those of subhalos.  But that work focused on 
a search for a particular benchmark model (dark matter 
with a mass of $40~\gev$, annihilating to $b \bar b$), 
and bounds from searches of dSphs can be significantly 
improved if the energy information of the photon 
spectrum is included.  In fact, for the scenario in 
which the blazar distribution is known, the normalization 
for the dark matter signal required for detection in 
the analysis of Ref.~\cite{Somalwar:2020awt} is about a 
factor of $\sim 5$ larger than the normalization we consider here (see~\cite{Boddy:2020}).  This 
suggests that more realistic modeling of both Poisson and non-Poisson backgrounds could impact our ability to extract information regarding dark matter microphysics.  
But echoing the conclusion of Ref.~\cite{Somalwar:2020awt}, our results at 
least suggest that an application of this 
technique to the actual Fermi data could be fruitful, 
especially considering the large uncertainties associated with all indirect dark matter searches. }

{\bf Acknowledgements}

JK is supported in part by DOE grant DE-SC0010504.
JR is supported by NSF grant AST-1934744.
\renew{The technical support and advanced computing resources from University of Hawai‘i Information Technology Services – Cyberinfrastructure, funded in part by the National Science Foundation MRI award \#1920304, are gratefully acknowledged.}


\bibliography{thebib}

\end{document}